\documentclass[lettersize,journal]{IEEEtran}
\usepackage{graphicx}
\usepackage{easyReview}
\usepackage{verbatim}
\usepackage{epstopdf}
\usepackage{setspace}
\usepackage{mathtools} 
\usepackage{mdframed} 
\usepackage{booktabs}
\usepackage{multirow}
\usepackage[most]{tcolorbox}
\usepackage{amsmath}
\usepackage{amsthm}
\usepackage{mathrsfs}
\usepackage{extarrows}
\usepackage{cite}
\newcommand{\orcidlink}[1]{%
  \href{https://orcid.org/#1}{%
    \tikz[baseline=-0.55ex]\node[
      fill=lime,
      circle,
      inner sep=0.35pt,
      text=white,
      font=\sffamily\bfseries\tiny
    ]{iD};%
  }%
}

\newtheorem{theorem}{Theorem}

\usepackage[bookmarks=false,colorlinks=true,linkcolor=blue!60!black,citecolor=blue!60!black,urlcolor=blue!60!black]{hyperref}
\usepackage{cleveref}
\usepackage{amssymb}
\newtheorem{remark}{Remark}

\newcommand{\RNum}[1]{\uppercase\expandafter{\romannumeral #1\relax}}
\newtheorem{lemma}{Lemma}

\newtheorem{problem}{Problem}
\newtheorem{definition}{Definition}

\newtheorem{corollary}{Corollary}

\usepackage{stfloats}
\usepackage{color, soul}
\usepackage{tikz,xcolor}
\usetikzlibrary{arrows.meta,automata, positioning, backgrounds, matrix, calc}

\newcommand{\sampleaction}{\textsf{sample}}
\newcommand{\preemptaction}{\textsf{preempt}}

\newtcbox{\hiRatio}{on line,
  colback=white, colframe=blue!70!black,
  boxrule=0.6pt, arc=1.5pt,
  left=1pt,right=1pt,top=0.5pt,bottom=0.5pt}
\definecolor{lime}{HTML}{A6CE39}

\makeatletter
\DeclareRobustCommand{\orcidicon}{%
	\begin{tikzpicture}
		\draw[lime, fill=lime] (0,0) 
		circle [radius=0.16] 
		node[white] {{\fontfamily{qag}\selectfont \tiny ID}};    \draw[white, fill=white] (-0.0625,0.095) 
		circle [radius=0.007];    \end{tikzpicture}
	\hspace{-2mm}}
\foreach \x in {A, ..., Z}{%
	\expandafter\xdef\csname orcid\x\endcsname{\noexpand\href{https://orcid.org/\csname orcidauthor\x\endcsname}{\noexpand\orcidicon}}
}

\newcommand*\bigcdot{\mathpalette\bigcdot@{.5}}
\newcommand*\bigcdot@[2]{\mathbin{\vcenter{\hbox{\scalebox{#2}{$\m@th#1\bullet$}}}}}
\makeatother

\usepackage[linesnumbered,ruled,vlined]{algorithm2e}

	\IEEEoverridecommandlockouts
	\makeatletter
	\def\@thmnote#1{\textit{[#1]}} 
	\makeatother

\begin{document}
		\title{{\fontsize{25pt}{29pt}\selectfont Optimal Preemption: Guaranteeing Information Freshness in Mean Age and Beyond}}
		\author{
			Aimin Li \orcidlink{0000-0003-3687-4378}, \emph{Member, IEEE}, Yiğit İnce \orcidlink{0009-0003-0465-0297}, \emph{Student Member, IEEE}, and
			 Elif Uysal \orcidlink{0000-0002-7258-4872}, \emph{Fellow, IEEE}
   \thanks{An earlier version of this paper was
presented at the IEEE International Symposium on Information Theory
(ISIT) 2026 \cite{aiminpreemption}. This work was supported in part by the European Union (through ERC Advanced
Grant 101122990-GO SPACE-ERC-2023-AdG), by the Scientific and Technological Research Council of Türkiye (TÜBİTAK) 1515 Frontier R\&D Laboratories Support Program for Türk Telekom 6G R\&D Lab under project number 5249902, and by TÜBİTAK 2247-A grant 123C572. Views and opinions expressed
are those of the authors only and do not necessarily reflect
those of the funding agencies. A. Li and Y. İnce contributed equally to this work. (Corresponding author: Elif Uysal.)

A. Li was with the Communication Networks Research Group (CNG), Department of Electrical and Electronics Engineering, METU, 06800 Ankara, Türkiye (e-mail: aimin@metu.edu.tr). He is now with the Division of Information Science and Engineering, KTH Royal Institute of Technology, 10044 Stockholm, Sweden.

Y. İnce and E. Uysal are with the Communication Networks Research Group (CNG), Department of Electrical
and Electronics Engineering, Middle East Technical University (METU),
06800 Ankara, Türkiye (e-mail: yigit.ince@metu.edu.tr; uelif@metu.edu.tr).}	
			}

\markboth{}%
{Li \MakeLowercase{\textit{et al.}}: Optimal Preemption}
\maketitle
		\allowdisplaybreaks

\begin{abstract}
Once a status update enters service, age-optimal policies must wait for it, even when heavy-tailed delays inflate the Age of Information (AoI). We develop a framework in which the sender may preempt the update in service and sample a fresh one, under general service-time distributions and explicit action costs. Our framework covers sources that generate updates at will as well as random update arrivals. Under nondecreasing age costs, we prove that, with preemption, a finite mean delay suffices for a finite average age even under infinite delay variance. We show the optimality of constant service-age thresholds from prior work for exponential service, and exhibit numerically that for general delays the optimal preemption rule depends on both the AoI and the elapsed service time. Under an action-cost budget, we further minimize the Value at Risk (VaR), which bounds the AoI except for a prescribed fraction of time, and the Conditional Value at Risk (CVaR), the mean AoI over that fraction. We reduce both to the same optimality equations without enlarging the state space. Under Pareto and log-normal delays, our algorithm \textsc{Tailor} keeps the average cost bounded as delay variance grows, while optimal non-preemptive sampling diverges. With optimal preemption, a more variable network can allow fresher data than a steadier one of equal mean delay.
\end{abstract}

\begin{IEEEkeywords}
Age of information, preemption, optimal sampling, impulse control, heavy-tailed delay, value at risk, conditional value at risk.
\end{IEEEkeywords}
		
		\begin{table*}[t]
\centering
\caption{Comparison with Representative Frameworks}
\label{tab:model_comparison}
\renewcommand{\arraystretch}{1.08}
\setlength{\tabcolsep}{7pt}
\small
\begin{tabular}{llccc}
\toprule
\textbf{Literature}
& \textbf{Arrival Model}
& \textbf{Generation Epochs}
& \textbf{Preemption Policy}
& \textbf{Tail-Risk Objective} \\
\midrule
Sun \emph{et al.} \cite{sun2017update}
& Generate-at-will
& Optimal
& $\times$
& $\times$
\\
Arafa \emph{et al.} \cite{arafa2019preemption}
& Generate-at-will
& Optimized
& Constant cutoff (heuristic)
& $\times$
\\
Akar \emph{et al.} \cite{akar2026squeezingpreemption}
& Poisson arrivals
& Exogenous
& Constant cutoff (heuristic)
& $\times$
\\
Moltafet \emph{et al.} \cite{moltafet2025probabilistic}
& Poisson arrivals
& Exogenous
& Fixed probability (heuristic)
& $\times$
\\
\midrule
\multirow{2}{*}{\textbf{This Work}}
& Poisson arrivals
& Exogenous
& \textbf{Causally optimal}
& \textbf{VaR, CVaR}
\\
& Generate-at-will
& \textbf{Jointly optimal}
& \textbf{Jointly optimal}
& \textbf{VaR, CVaR}
\\
\bottomrule
\end{tabular}
\end{table*}

\section{Introduction}\label{sectionI}
		\IEEEPARstart{A}{ge} of Information (AoI) has emerged as a fundamental performance metric for real-time sensing, communication, and control systems \cite{yates2021age,kosta2017age,chen2025aoioffloading,liu2023aos,xie2024multisensor}, shifting the design paradigm from how much data is delivered (throughput-oriented) to how fresh and effective the delivered information is (goal-oriented) \cite{uysal2022semantic,gunduz2023timely,li-goal,10562359, uysal_li_goal}. In practice, networked systems often experience highly variable, heavy-tailed delays caused by buffering, routing, or retransmissions~\cite{xie2023importance}. To mitigate such uncertainty, a rich body of literature (e.g., \cite{sun2017update,sun2019wiener,TangCWYT23,bedewy2021optimal,li2024sampling, 11393542,liyanaarachchi2024role,10619298,atasayar2025fresh,chen2024optimal,pan2023sampling,li2025optimal,10559951}) has established optimal sampling policies without preemption. A key takeaway is that age-optimal sampling under heavy-tailed delays often \emph{introduces deliberate nonzero waiting}, leading to the \emph{counterintuitive} principle that ``\emph{lazy is timely}.''

\begin{figure}[!t]
\centering
\includegraphics[width=0.8\columnwidth]{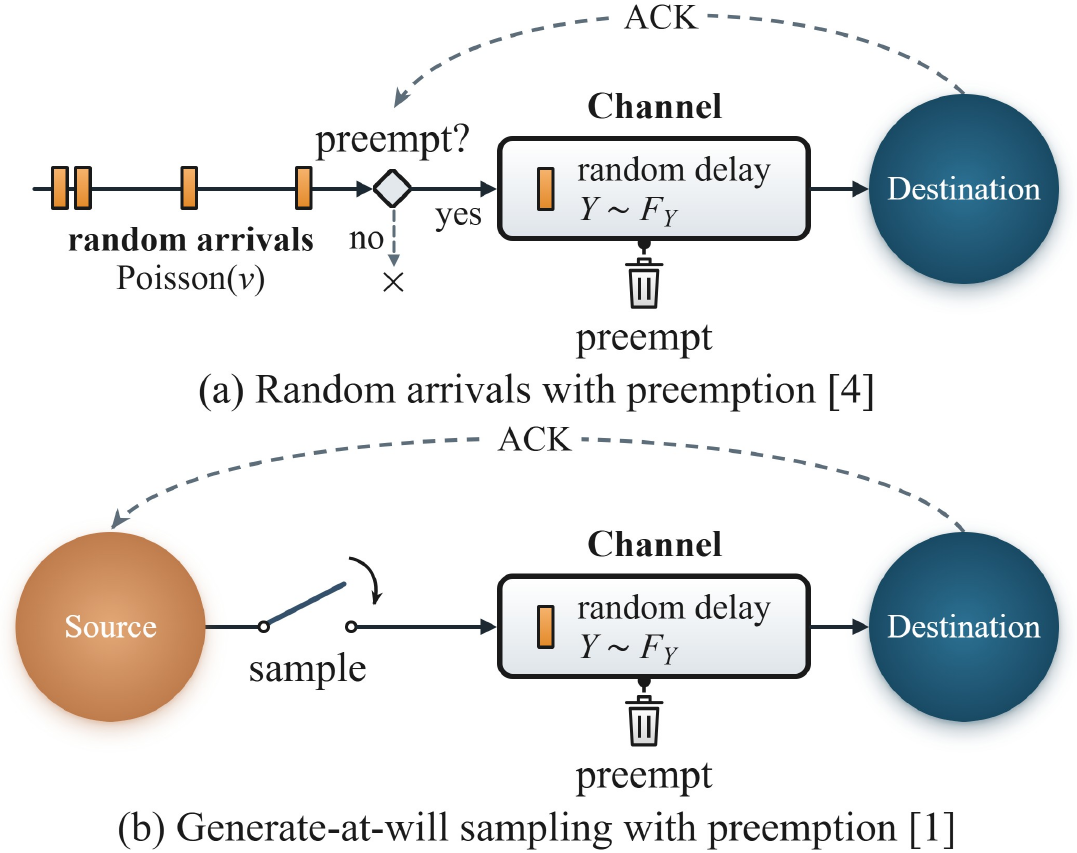}
\caption{System model. Both paradigms share a bufferless channel with service time $Y\sim F_Y$ and instantaneous ACKs, differing only at the input. (a) Under random arrivals \cite{akar2026squeezingpreemption}, the controller decides at each busy arrival whether to persist with the update in service or preempt it. (b) Under generate-at-will, the controller decides when to sample \cite{sun2017update} and when to preempt \cite{aiminpreemption}; sampling turns the exogenous arrival process of (a) into a control variable.}
\label{fig:model}
\end{figure}

        While effective prior to transmission, we argue that \emph{lazy sampling alone suffers from post-transmission blindness}. Under the prevailing non-preemptive assumption, once a packet enters service, the controller becomes a \textit{passive bystander} until an acknowledgment arrives. This passivity induces severe \textit{head-of-line} blocking, preventing fresher updates from being transmitted. The effect is particularly harmful under heavy-tailed delays (\emph{e.g.}, Pareto or log-normal), which may exhibit the Increasing Mean Residual Life (IMRL) property \cite{lawless2011statistical}: the longer a packet has already spent in service, the longer it is expected to remain there. In such regimes, sunk service time signals prolonged stagnation, not imminent completion.
        
\begin{figure}[!t]
        \centering
        
        \newlength{\figpairh}\settoheight{\figpairh}{\includegraphics[width=0.82\columnwidth]{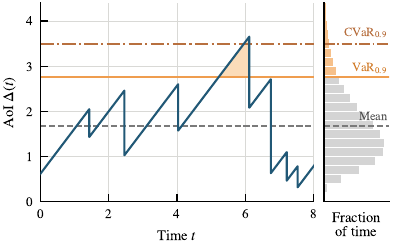}}%
        \includegraphics[width=\columnwidth,height=\figpairh]{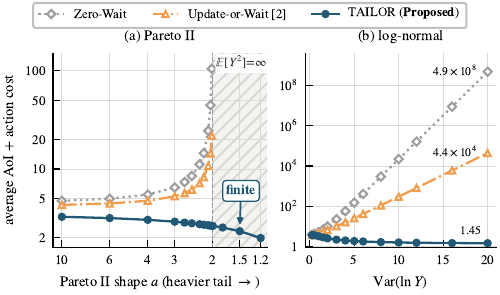}
\caption{Preemption turns delay variability into an advantage. The average cost $\bar\Delta_\pi+\lambda\bar K_\pi$ adds the average AoI and the average action cost. (a) Pareto~II service, with the tail heavier to the right: the average cost of every non-preemptive policy diverges as the shape $a$ decreases to $2$, whereas \textsc{Tailor} remains finite; the hatched region marks $a\le2$, where $\mathbb{E}[Y^2]=\infty$. (b) Log-normal service: the same costs grow exponentially in $\operatorname{Var}(\ln Y)$, whereas the cost of \textsc{Tailor} decreases. Both panels use $\mathbb{E}[Y]=2$ and $\kappa_s=\kappa_p=1$; Zero-Wait and Update-or-Wait~\cite{sun2017update} do not preempt.}
        \label{fig:Preemption}
\end{figure}

        Overcoming this stagnation requires unlocking \emph{preemption} as an active control lever by aborting an aging in-service packet to make way for a fresh update, as illustrated in Fig.~\ref{fig:model}. By actively terminating a stalled transmission, preemption breaks head-of-line blocking and truncates the heavy tail of residual service delays. Fig.~\ref{fig:Preemption} previews the payoff: as the service-time variance grows, the non-preemptive AoI-optimal policy degrades, whereas optimal preemption keeps lowering the average cost. However, this control lever introduces a nontrivial trade-off: ($i$) \textit{persisting} with the ongoing packet preserves the service effort already invested, but risks incurring a long remaining delay, whereas ($ii$) \textit{preempting} sacrifices prior service progress and incurs a fixed switching cost, but truncates the heavy tail of future delays. In other words, the controller must weigh the expected marginal benefit of continuing the current, already-aged service attempt against the cost and renewal benefit of restarting with a fresh sample.

        
        Deploying preemption requires no new system primitives, because the underlying cancellation hooks are already available across modern infrastructures. In 3GPP 5G New Radio (NR), the link layer discards a service data unit (SDU) once its discard timer expires~\cite{3gpp38323,3gpp38322}. Similarly, edge computing and robotic runtimes expose interfaces to abort ongoing tasks via remote procedure call (RPC) cancellation \cite{grpcCancellation}, inference request preemption \cite{vllmPreemption}, and goal cancellation in the Robot Operating System (ROS) \cite{ros2Actions}. Yet in current deployments, these hooks are governed by static expiration timers or ad hoc resource heuristics, leaving open when a constant timeout is theoretically justified and when state-dependent rules are required. What is missing is a principled rule for firing these hooks, which raises a question that spans communication and computation: \emph{When should a system abort an ongoing, aging update to restart with fresher information?}

\begin{figure}[!t]
\centering
\includegraphics[width=0.75\columnwidth]{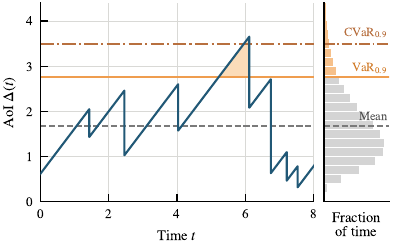}
\caption{Mean, VaR, and CVaR of the AoI under Zero-Wait sampling without preemption and log-normal service times. Left: a sample path, with the AoI above $\operatorname{VaR}_{0.9}$ shaded. Right: the long-run fraction of time that the AoI spends at each level. $\operatorname{VaR}_{0.9}$ is the AoI level exceeded during $10\%$ of the time, and $\operatorname{CVaR}_{0.9}$ is the mean AoI over that time; both characterize the upper tail of the distribution, which the mean does not reveal.}
\label{fig:VaR_CVaR_demo}
\end{figure}

Furthermore, heavy-tailed service times not only elevate average AoI, but also expose mission-critical systems to rare yet severe episodes of information staleness. A policy tailored solely to mean AoI remains blind to distributional tails, motivating the joint consideration of average freshness and tail-risk metrics, specifically the Value at Risk (VaR) and Conditional Value at Risk (CVaR) of the AoI distribution \cite{rockafellar2000cvar}. Fig.~\ref{fig:VaR_CVaR_demo} illustrates the difference on a sample path: a single update stuck in service drives the AoI far above its mean, an excursion that the mean dilutes but VaR and CVaR expose. However, optimizing non-linear distributional metrics over continuous-time hybrid dynamics typically requires augmenting the state space to track accumulated risk, which causes a \emph{curse of dimensionality} over continuous age and service-time states.

To bridge these theoretical and algorithmic gaps, we establish a unified continuous-time optimal control framework for joint sampling and preemption under general service-time distributions and explicit operational penalties. We investigate two arrival paradigms, illustrated in Fig.~\ref{fig:model}: an exogenous random-arrival model \cite{akar2026squeezingpreemption,moltafet2025probabilistic} (where the controller decides whether an arriving update preempts the one in service) and an endogenous generate-at-will (GaW) model, in which the controller decides both sampling and preemption epochs.

        Our main contributions are fourfold:
\begin{itemize}
    \item \textbf{System Model:} 
    Prior models (Table~\ref{tab:model_comparison}) sample without preemption, preempt only in slotted time with bounded delays, or target only the mean AoI under a single arrival paradigm. We remove these restrictions: our controller preempts at any time under general delays and action costs, under both arrival paradigms, for general running costs, the VaR, and the CVaR of the AoI (see Fig.~\ref{fig:VaR_CVaR_demo}).

\item \textbf{Optimality Structure:} 
    Modeling the system as a piecewise-deterministic Markov process, we derive average-cost optimality equations for both arrival paradigms and, under GaW, reduce preemption to optimal stopping against a random completion time. Under exponential service, we prove that the constant service-age thresholds analyzed in~\cite{arafa2019preemption,akar2026squeezingpreemption} are optimal (Theorem~\ref{thm:exp}), whereas general delays induce AoI-dependent boundaries. While non-preemptive sampling~\cite{sun2017update} requires $\mathbb{E}[Y^2]<\infty$, preemption keeps the AoI finite if $\mathbb{E}[Y]<\infty$ (see Fig.~\ref{fig:Preemption}(a)).

    \item \textbf{Tail-Risk Optimization Framework:} 
    Freshness service-level agreements constrain the tail of the stationary AoI distribution, which lacks the additivity required by dynamic programming. Prior work restores additivity for the CVaR of discounted accumulated AoI by augmenting the state~\cite{zhou2020risk}. Without augmentation, we develop a unified framework that minimizes the stationary VaR and CVaR under an action-cost budget: a single threshold turns each into an average cost, and a one-dimensional search with performance guarantee recovers the exact metric.
    \item \textbf{Algorithmic Design and Insights:}
    We develop \textsc{Tailor} (\textbf{TAIL}-aware \textbf{O}ptimal p\textbf{R}eemption), a policy-iteration algorithm for the optimality equations over unbounded state spaces. Our simulations (Section~\ref{sec:simulations}) show that \textsc{Tailor} keeps the average cost bounded under heavy-tailed delays where optimal non-preemptive sampling~\cite{sun2017update} diverges, and that AoI-dependent preemption gains little over Static Preemption~\cite{arafa2019preemption,akar2026squeezingpreemption} for the average cost but substantially lowers the CVaR. \emph{Optimal preemption thus turns delay variability from an operational liability into a strategic asset for information freshness.}
\end{itemize}

      \section{Related Work}
\textbf{Non-Preemptive Sampling and Prescribed Preemption.}
Sun \emph{et al.} \cite{sun2017update} established age-optimal sampling under random delay and showed that waiting before sampling can reduce AoI. Subsequent works extended this approach to multi-source and goal-oriented settings \cite{bedewy2021optimal,li2024sampling}. These policies act only before transmission: once an update enters service, the controller cannot react to its elapsed service time, and a finite average AoI requires $\mathbb{E}[Y^2]<\infty$. Queueing studies, in contrast, fix the service discipline \emph{a priori} and evaluate the resulting AoI under FCFS, LCFS, packet management, and priority rules, including energy-harvesting settings \cite{kaul2012status,costa2016age,chen2016age,najm2017status,kaul2018age,soysal2021age,farazi2018harvesting_preemption}. Their results show that preemption is not universally beneficial: under gamma-distributed service, LCFS with preemption underperforms its non-preemptive counterpart as the service time becomes less variable \cite{Gammaawakening}. Under non-exponential delays, the residual service time depends on the service age, while the gain from restarting depends on the monitor AoI; preemption should therefore depend on both states.

\textbf{Preemption as a Control Decision.}
Treating preemption as a decision variable, discrete-time studies exploit packet-level side information \cite{wang2019preemption} or formulate slotted MDPs with bounded delays and no action costs \cite{banerjee_ulukus_isit24}. In continuous time, prior works optimize within parameterized policy classes. Arafa \emph{et al.} \cite{arafa2019preemption} pair a waiting threshold with a fixed preemption cutoff for GaW updates, without action costs. Champati \emph{et al.}~\cite{Champati2021minimum} prove that a fixed threshold minimizes peak AoI under i.i.d. service and request delays. Akar \emph{et al.} \cite{akar2026squeezingpreemption} tune a constant service-age threshold or preemption probability under Poisson arrivals without using the monitor AoI. Moltafet \emph{et al.} \cite{moltafet2025probabilistic} derive the AoI of M/G/1/1 queues in which every busy arrival preempts with a fixed probability, and a recent extension~\cite{moltafet2026elapsed} lets it vary with the elapsed service time in multi-source queues, still independently of the monitor AoI. Along the same line, \cite{asvadi2025delayed} optimizes a static delayed-preemption deadline in prioritized networks, and \cite{erbayat2025corr_preempt_aoi} adopts a fixed preemption probability for correlated sources under exponential service. Whether these AoI-independent rules remain optimal for average AoI with action costs, and what replaces them under general delays, remains open. As summarized in Table~\ref{tab:model_comparison}, we optimize over all admissible causal policies in continuous time, under both arrival paradigms and with explicit sampling and preemption costs. We prove that the static thresholds postulated in \cite{arafa2019preemption,akar2026squeezingpreemption} are optimal under exponential service, and our numerics show that general delays induce preemption boundaries depending jointly on AoI and service age.

\textbf{Tail risk of AoI.}
The AoI violation probability $\mathbb P(\Delta>x)$ is a widely studied tail measure of freshness. Analytical studies derive stationary AoI distributions and violation bounds for single-server queues~\cite{Inoue2019general,Champati2019distribution}, the outage probability of the peak AoI~\cite{Seo2019outage}, and risk-aware and extreme-value tail metrics~\cite{xiao2024statistical,miridakis2022extreme}. Design studies tune the update rate to satisfy a violation-probability target~\cite{Champati2021statistical}, provision short-packet transmission under peak-age or statistical QoS guarantees~\cite{Devassy2019reliable,chen2025rsmaxurllc,zhang2025fbl}, control tail behavior using Lyapunov optimization and extreme-value theory~\cite{Liu2019taming,AbdelAziz2020optimized}, or assign tasks according to a conditional age-at-risk metric~\cite{huang2026conditional}. These studies establish the importance of tail-aware AoI design but leave a key structural difficulty open: the VaR and the CVaR are a quantile and a tail expectation of the stationary AoI distribution rather than long-run averages of a running cost, so they lack the additive structure of standard dynamic programming. Existing approaches address this difficulty only indirectly. Zhou \emph{et al.}~\cite{zhou2020risk} recover additivity by augmenting the state with auxiliary risk levels, which yields history-dependent policies and a CVaR of the discounted accumulated AoI rather than of the stationary distribution. Ling \emph{et al.}~\cite{ling2022age} optimize the average AoI and use the CVaR only to approximate distributionally robust energy and outage constraints under imperfect channel state information. Learning-based approaches rely on sample approximations. Pan \emph{et al.} control the return-based CVaR of the version AoI with distributional reinforcement learning~\cite{pan2026tailrisk} and penalize consecutive age violations from sampled returns~\cite{pan2026persistence}. By contrast, we design an optimization framework that resolves this difficulty without state augmentation and minimizes the VaR and the CVaR of the stationary AoI distribution over state-dependent sampling and preemption policies. 

Finally, this paper extends our previous work \cite{aiminpreemption}, which studied mean-AoI minimization under GaW sampling with $\mathbb{E}[Y^2]<\infty$, in four directions: $i$) we incorporate exogenous random arrivals, where preemption is restricted to arrival epochs, thereby covering sources whose update generation cannot be controlled; $ii$) we generalize the objective from the AoI itself to nondecreasing AoI cost functions, including the indicator and hinge costs underlying VaR and CVaR; $iii$) we relax the service-time assumption to $\mathbb{E}[Y]<\infty$, allowing $\mathbb{E}[Y^2]=\infty$ and thus covering heavy-tailed regimes in which non-preemptive policies can incur infinite mean AoI; and $iv$) we develop a budget-constrained framework for optimizing the VaR and CVaR of the AoI, which controls the tail risk rather than only the mean.

\section{System Model}
\label{sec:model}
We consider a single-source status update system, as depicted in Fig.~\ref{fig:model}. A source sends time-stamped updates to a destination over a channel with random service time, and a controller at the source decides whether to persist with or preempt the update in service. We study two update-generation mechanisms, exogenous random arrivals and GaW sampling, described in Section~\ref{subsec:arrival_models}.

\subsection{Channel Model}
\textbf{Service Delay.} The server holds at most one update, and each update is generated at the instant it enters service, namely upon its arrival or upon sampling. Let $Y_k$ denote the potential service time of the $k$-th service attempt, i.e., the time it needs to complete if it is not interrupted. We assume that $\{Y_k\}$ are independent and identically distributed (i.i.d.) copies of a nonnegative random variable $Y$, independent of the arrival process and of past decisions, with an absolutely continuous cumulative distribution function (CDF) $F_Y(\cdot)$, density $f_Y(\cdot)$, and tail $\bar F_Y(t)\triangleq\Pr\{Y>t\}$. We assume $\mathbb{E}[Y]<\infty$ and impose no condition on higher moments. In particular, $Y$ may have infinite variance. A preempted update is cancelled immediately and is never delivered, and the next attempt starts without delay. The controller observes the elapsed service time and whether the attempt has completed, but not $Y_k$.

\textbf{System State and Dynamics.} Let $U(t)$ be the generation time of the most recently delivered update and $S(t)$ the start time of the update in service. The AoI at the destination and the service age are
\begin{equation}
\Delta(t)\triangleq t-U(t),\qquad b(t)\triangleq t-S(t),
\end{equation}
with $b(t)\triangleq 0$ when the server is idle. The system state is the hybrid process $X(t)=(\Delta(t),b(t),m(t))$, where $m(t)\in\{I,B\}$ is the channel mode (idle or busy). Its paths are right-continuous with left limits and lie in the reachable set
\begin{equation}
\mathcal S\triangleq\{(\Delta,0,I):\Delta\ge 0\}\cup\{(\Delta,b,B):0\le b\le\Delta\},
\end{equation}
starting from an idle state $X(0)=(\Delta_0,0,I)$. Between jumps, the state evolves deterministically as $\dot\Delta(t)=1$ and $\dot b(t)=\mathbf 1_{\{m(t)=B\}}$; that is, without an action the system simply keeps waiting or serving. Three jump maps on $\mathcal S$ describe all transitions of the state:
\begin{subequations}\label{eq:jumps}
\begin{align}
\mathcal M_c(\Delta,b,B)&=(b,0,I),\label{eq:Mc}\\
\mathcal M_s(\Delta,0,I)&=(\Delta,0,B),\label{eq:Ms}\\
\mathcal M_p(\Delta,b,B)&=(\Delta,0,B).\label{eq:Mp}
\end{align}
\end{subequations}
A \emph{service completion} occurs at the hazard rate $\gamma(b)$ defined in \eqref{eq:hazard}. Since the delivered update was generated when its service started, the AoI drops to the service age $b$, as in \eqref{eq:Mc}. Starting service in the idle mode switches the mode to busy and leaves the AoI unchanged, as in \eqref{eq:Ms}. \emph{Preemption} cancels the update in service and immediately starts a fresh one, resetting the service age while leaving the AoI unchanged, as in \eqref{eq:Mp}. If a completion and a control action coincide, the completion is processed first and the action applies to the post-completion state, so sampling may follow a delivery immediately.

\textbf{Residual Life and Hazard Rate.} At service age $b$, the controller weighs persisting against preempting. Persisting preserves the service effort already invested but risks a long remaining delay, whereas preempting pays $\kappa_p$ and discards this effort to restart with a fresh update whose service time is a new draw of $Y$. The decision thus hinges on the conditional law of the remaining service time given $b$.
\begin{definition}[Residual Life and Hazard Rate \cite{shortle2018fundamentals}]
For a service age $b\ge 0$ with $\bar F_Y(b)>0$, the \emph{residual service time} is $R_b\triangleq (Y-b\mid Y>b)$, with CDF and survival function
\begin{equation}\label{eq:residual_cdf}
\begin{aligned}
F_{R_b}(t)&=\frac{F_Y(b+t)-F_Y(b)}{\bar F_Y(b)},\\
S_b(t)&\triangleq\Pr\{R_b>t\}=\frac{\bar F_Y(b+t)}{\bar F_Y(b)},\qquad t\ge 0.
\end{aligned}
\end{equation}
The \emph{hazard rate}, i.e., the instantaneous conditional completion rate at service age $b$, is
\begin{equation}\label{eq:hazard}
\gamma(b)\triangleq\lim_{\delta\downarrow 0}\frac{\Pr\{Y\le b+\delta\mid Y>b\}}{\delta}=\frac{f_Y(b)}{\bar F_Y(b)}.
\end{equation}
\end{definition}
The hazard rate captures how the service age shifts this trade-off. Under an increasing failure rate (IFR), i.e., a nondecreasing $\gamma(b)$, an update becomes more likely to complete as it ages, so continued service grows more promising. For exponential $Y$, $\gamma(b)$ is constant, so the service age carries no information about the remaining service time, although it still determines the age of the update upon delivery. Heavy-tailed delays reverse the IFR behavior: the Pareto distribution has a decreasing failure rate (DFR), and the log-normal hazard rate eventually decreases. A lingering update then becomes ever less likely to finish soon, which favors preemption.

\textbf{Running Cost.} The running cost $g:\mathbb R_+\to\mathbb R_+$ is nondecreasing and satisfies the polynomial growth condition $g(\Delta)\le C(1+\Delta^{\ell})$ for all $\Delta\ge 0$, for some constant $C>0$ and integer $\ell\ge 0$. This class includes the linear AoI and the tail-risk running costs in Section~\ref{subsec:risk_formulations}.

\subsection{Arrival Models}
\label{subsec:arrival_models}
The two paradigms share the dynamics above but differ in the controller's decisions. In the \emph{random-arrival} model, updates arrive exogenously, and the controller decides at each busy arrival whether to preempt. In the \emph{GaW} model, it decides both when to sample and when to preempt.

\subsubsection{Random Arrivals}
Updates arrive according to a Poisson process with rate $\nu>0$, independent of $\{Y_k\}$, at epochs $t_1<t_2<\cdots$. This model follows the single-source preemption setting of \cite{akar2026squeezingpreemption}, and its AoI under fixed LCFS disciplines with and without preemption is analyzed in \cite{Gammaawakening,najm2017status}.
\begin{itemize}
\item \textbf{Transitions.} An arrival in the idle mode immediately enters service via $\mathcal M_s$. An arrival in the busy mode, with pre-arrival state $X(t_i^-)=(\Delta,b,B)$, is either discarded, which leaves the state unchanged, or preempts the update in service via $\mathcal M_p$.
\item \textbf{Action Costs.} Each idle admission incurs an admission cost $\kappa_s\ge 0$, and each preemption incurs a cost $\kappa_p\ge 0$ that covers cancelling and restarting; discarding is free.
\item \textbf{Admissible Policy Space.} Let $\{\mathcal F_t\}$ be the observation filtration generated by the initial state, the arrival and completion epochs, and the past decisions, augmented by an independent randomization variable for randomized policies. Because a discarded arrival does not change $X$, this filtration is richer than that of $X$. An admissible policy $\pi\in\Pi_R$ is a sequence of $\mathcal F_{t_i}$-measurable decisions $a_i\in\{\textsf{discard},\preemptaction\}$ at busy arrivals.
\end{itemize}

\subsubsection{Generate-at-Will}
Updates are generated on demand, and we model the system as an impulse-controlled piecewise-deterministic Markov process (PDMP) \cite{davis1984piecewise}.
\begin{itemize}
\item \textbf{Impulses and Costs.} The admissible action set is mode dependent: $\mathcal A(\Delta,b,I)=\{\sampleaction\}$ and $\mathcal A(\Delta,b,B)=\{\preemptaction\}$. Action $\sampleaction$ generates a fresh update that immediately enters service via $\mathcal M_s$, and action $\preemptaction$ cancels the update in service and restarts with a fresh one via $\mathcal M_p$. Each impulse incurs a fixed cost, where $\kappa_p$ covers both the cancellation and the new update:
\begin{equation}\label{eq:impulse_cost}
K(x,\sampleaction)=\kappa_s\ge0,\ \ K(x,\preemptaction)=\kappa_p\ge0.
\end{equation}
\item \textbf{Admissible Policy Space.} Let $\{\mathcal F_t\}$ be the observation filtration generated by the initial state, the completion epochs, and the past impulses, augmented by an independent randomization variable when randomized policies are considered. An admissible policy $\pi\in\Pi_G$ is a sequence $\{(\tau_i,a_i)\}_{i\ge 1}$ of impulse times and actions such that (i) each $\tau_i$ is an $\{\mathcal F_t\}$-stopping time and $a_i\in\mathcal A(\hat X(\tau_i))$ is $\mathcal F_{\tau_i}$-measurable, where $\hat X(\tau_i)$ is the state at $\tau_i$ after any completion at that instant and before the impulse; (ii) $\tau_i<\tau_{i+1}$ for all $i$; and (iii) $\tau_i\uparrow\infty$ a.s., so that impulses do not accumulate in finite time.
\end{itemize}

\section{Average-Cost Optimal Preemption}
\label{sec:age_optimal_preemption}
This section formulates a unified optimization problem that minimizes the long-term average age-related cost and characterizes the associated optimal control policies under the random-arrival and GaW models. We then develop \textsc{Tailor}, an algorithm that computes the optimal policy.

\subsection{Problem Formulation}
For a policy $\pi\in\Pi_R$ or $\pi\in\Pi_G$, define the long-run average running cost as
\begin{equation}
\bar{g}_\pi \triangleq
\limsup_{T\to\infty}\frac{1}{T}\,
\mathbb E^\pi\!\left[\int_0^T g(\Delta(t))\,dt
\right],
\end{equation}
and the long-run average action cost as
\begin{equation}
\label{eq:Kbar}
\bar{K}_\pi \triangleq \limsup_{T\to\infty}\frac{1}{T}\,
\mathbb{E}^\pi\!\left[\kappa_s N_s^\pi(T)+\kappa_p N_p^\pi(T)\right],
\end{equation}
where $N_s^\pi(T)$ counts the updates that enter service from the idle state and $N_p^\pi(T)$ counts the preemptions during $[0,T]$. Under $\Pi_G$, these are exactly the $\sampleaction$ and $\preemptaction$ impulses, so by \eqref{eq:impulse_cost} the bracket equals $\sum_{i:\tau_i\le T}K(\hat X(\tau_i),a_i)$. We restrict attention to policies for which the limits defining $\bar g_\pi$ and $\bar K_\pi$ exist, with $\bar K_\pi<\infty$, so that the weighted average cost $\rho^\pi\triangleq\bar{g}_\pi+\lambda\bar{K}_\pi$ coincides with the long-run average of the total cost, and we seek policies that minimize $\rho^\pi$. Sections~\ref{sec:age_optimal_preemption} and~\ref{sec:tail} optimize over this class; the stationary policies that Theorems~\ref{thm:random_integral_bellman} and~\ref{thm:integral_bellman} characterize have convergent averages and therefore belong to it.
\begin{problem}[Average Age-Cost Minimization]\label{p1}
\begin{equation}
\rho_i(g,\lambda)=\inf_{\pi\in\Pi_i}\left(\bar{g}_{\pi}+\lambda \bar{K}_\pi\right), \quad i\in \{R,\,G\}.
\end{equation}
\end{problem}
Here, $\lambda\ge 0$ weights the average action cost. In Section~\ref{subsec:kernel}, $\lambda$ serves as a Lagrange multiplier, which reduces the constrained VaR and CVaR problems to instances of Problem~\ref{p1}. When no ambiguity arises, we write $\rho$ for $\rho_i(g,\lambda)$.

\textbf{Technical challenges.} The natural characterization of Problem~\ref{p1} is a Hamilton--Jacobi--Bellman quasi-variational inequality (HJB--QVI) \cite{bensoussan1975nouvelles}, which differentiates the relative value function along the flow. Under the average-cost criterion, however, this function grows without bound in the AoI and may be nonsmooth where the optimal action switches, so the differential formulation remains \textit{heuristic} \cite{aiminpreemption}. A further feature complicates the analysis: the value of persisting depends both on how stale the destination is and on how long the update has been in service, so the state $(\Delta,b)$ is two-dimensional and unbounded. We therefore adopt an integral dynamic programming principle \cite{stettner2022approximation}: writing the optimality equations between consecutive events yields coupled integral average-cost optimality equations (ACOEs) that require no derivative of the relative value function and no density of $F_Y$. Let $h(\Delta,b,m)$ denote the relative value function, and write $h_I(\Delta)\triangleq h(\Delta,0,I)$, $h_B(\Delta,b)\triangleq h(\Delta,b,B)$, and $v(y)\triangleq h_B(y,0)$ for its value at the busy-start boundary.

\subsection{Optimal Preemption Under Random Arrivals}
\label{subsec:ra_acoe}
Under random arrivals, the controller acts only at exogenous Poisson epochs. At each busy arrival, it either persists with the ongoing transmission to preserve the invested service effort, or preempts it to restart with a fresh update. The subsequent transition hinges on a race between the residual service life $R_b$ and the next arrival: an earlier completion delivers the update and leaves the server idle, whereas an earlier arrival triggers the next decision. Theorem~\ref{thm:random_integral_bellman} formalizes this competition through coupled integral ACOEs. Unlike \cite{akar2026squeezingpreemption}, which considers only constant service-age thresholds, the theorem characterizes the optimal policy over the entire admissible space.
\begin{theorem}[Integral ACOEs Under Random Arrivals]\label{thm:random_integral_bellman}
Suppose that a scalar $\rho$ and Borel functions $(h_I(\cdot),h_B(\cdot,\cdot))$ of polynomial growth, normalized by $h_B(0,0)=0$, satisfy the following coupled ACOEs.

\textbf{I. Idle phase:} For all $\Delta\ge 0$,
\begin{equation}
\label{eq:random_h_idle}
\begin{split}
h_I(\Delta) &= \int_0^\infty e^{-\nu t} \big( g(\Delta + t) - \rho \big) \, dt \\
&\quad + \int_0^\infty \nu e^{-\nu t} h_B(\Delta + t, 0) \, dt + \lambda\kappa_s.
\end{split}
\end{equation}

\textbf{II. Busy phase:} For all $0\le b\le\Delta$ with $\bar F_Y(b)>0$,
\begin{equation}
\label{eq:random_h_busy}
\begin{split}
h_B(\Delta, b) &= \int_0^\infty e^{-\nu t} S_b(t) (g(\Delta + t) - \rho) \, dt \\
&\quad + \int_{[0, \infty)} e^{-\nu t} h_I(b+t) \, dF_{R_b}(t) \\
&\quad + \int_0^\infty \nu e^{-\nu t} S_b(t) \min\big\{ h_B(\Delta+t, b+t), \\
&\qquad\qquad \lambda\kappa_p + h_B(\Delta+t, 0) \big\} \, dt.
\end{split}
\end{equation}

\textbf{Optimal preemption policy:} If the following rule has finite average cost and \eqref{eq:transversality} in the supplementary material holds, then $\rho$ is optimal and the rule is optimal among admissible $\pi\in\Pi_R$ with finite average cost: at a busy arrival $(\Delta,b)$, preempt if and only if
\begin{equation}
\label{eq:random_preempt_rule}
\Phi_1(\Delta, b)\triangleq h_B(\Delta, b) - h_B(\Delta, 0) - \lambda\kappa_p\ge 0.
\end{equation}
\end{theorem}
\begin{proof}
See Appendix~\ref{app:random_integral_bellman_proof} in the supplementary material.
\end{proof}

\subsection{Joint Sampling and Preemption Under Generate-at-Will}
\label{subsec:gaw_acoe}
\enlargethispage{\baselineskip}
Under GaW, the controller may sample or preempt at any instant, so the busy-phase decision is a continuous-time stopping rule over $(\Delta,b)$. From a busy-start state $(y,0,B)$, the trajectory before completion or preemption is $(\Delta(t),b(t))=(y+t,t)$, along which $\Delta(t)-b(t)\equiv y$. This invariance collapses the busy phase onto the scalar busy-start AoI $y$ \cite{aiminpreemption} and reduces preemption to an optimal stopping problem against the random completion time. Theorem~\ref{thm:integral_bellman} establishes the resulting coupled ACOEs. It extends \cite[Thm.~1]{aiminpreemption} from the linear AoI with unweighted action costs to a general running cost $g$ and $\lambda$-weighted action costs, which the tail-risk reductions of Section~\ref{sec:tail} require.

\begin{figure}[!t]
\centering
\includegraphics[width=0.80\columnwidth]{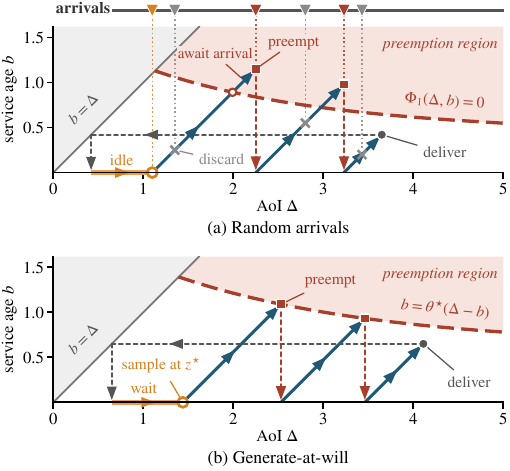}
\caption{One delivery cycle under the optimal policy for (a) random arrivals with $\nu=2$ and (b) GaW, with $g(\Delta)=\Delta$, log-normal service ($\mathbb{E}[Y]=2$, $\operatorname{Var}(\ln Y)=2$), and $\lambda=\kappa_s=\kappa_p=1$. Solid arrows are flows, dotted arrows are jumps, and the shaded area is the preemption region. The top rail in (a) marks arrival epochs, and both axes share one scale.}
\label{fig:trajectory}
\end{figure}

\begin{theorem}[Integral ACOEs and Optimal-Stopping Reduction Under Generate-at-Will]\label{thm:integral_bellman}
Suppose that a scalar $\rho$ and Borel functions $(h_I(\cdot),v(\cdot))$ of polynomial growth, normalized by $v(0)=0$, satisfy the following coupled ACOEs.

\textbf{I. Idle phase:} For all $\Delta\ge 0$,
\begin{equation}\label{eq:idle_integral_bellman}
\begin{aligned}
h_I(\Delta)&=\inf_{z\ge \Delta} Q_I(\Delta,z;\rho),\\
Q_I(\Delta,z;\rho)&\triangleq\int_{\Delta}^{z} g(t)\,dt-\rho(z-\Delta)+\lambda\kappa_s+v(z).
\end{aligned}
\end{equation}

\textbf{II. Busy phase:} For all $y\ge 0$,
\begin{equation}\label{eq:busy_integral_bellman}
v(y)=\inf_{\theta\ge 0} Q(y,\theta;\rho),
\end{equation}
where the busy stage with preemption age $\theta$ has relative cost
\begin{equation}\label{eq:busy_acoe_Q}
\begin{aligned}
Q(y,\theta;\rho)= &\int_{0}^{\theta} g(y+t) \overline{F}_Y(t) \, dt - \rho A(\theta)
\\&+\int_{0}^\theta h_I(t)\,dF_Y(t)
+\bar F_Y(\theta)\big(\lambda\kappa_p+v(y+\theta)\big),
\end{aligned}
\end{equation}
with $A(\theta)\triangleq \mathbb{E}[\min\{Y,\theta\}]=\int_0^\theta \bar F_Y(t)\,dt$.
If the infima in \eqref{eq:idle_integral_bellman} and \eqref{eq:busy_integral_bellman} are attained and \eqref{eq:transversality} in the supplementary material holds, any measurable selection of minimizers whose induced impulse times satisfy the admissibility conditions in Section~\ref{sec:model}, e.g., with $\theta^\star(y)\ge\epsilon$ for some $\epsilon>0$, induces a stationary policy that is optimal among all admissible $\pi\in\Pi_G$ with finite average cost, as follows.

\textbf{(a) Optimal sampling policy:} From an idle state $(\Delta,0,I)$, wait until the AoI reaches
\begin{equation}\label{eq:zstar_def}
z^\star(\Delta)\in \arg\min_{z\ge \Delta} Q_I(\Delta,z;\rho),
\end{equation}
that is, for $u^\star(\Delta)=z^\star(\Delta)-\Delta$ time units, and then sample.

\textbf{(b) Optimal preemption policy:} From a busy-start AoI $y$, continue service until it completes or until the service age reaches the threshold
\begin{equation}\label{eq:taustar_def}
\theta^\star(y)\in \arg\min_{\theta\ge 0} Q(y,\theta;\rho),
\end{equation}
whichever occurs first. In the latter case, preempt and restart with a fresh update, which starts a new service attempt with busy-start AoI $y+\theta^\star(y)$.
\end{theorem}
\begin{proof}[Proof sketch]
We follow \cite[App.~B]{aiminpreemption} with the running cost $\Delta$ replaced by $g(\Delta)$ and each impulse cost $\kappa$ replaced by $\lambda\kappa$. The reduction of any admissible rule to a deterministic sampling target $z$ and a deterministic stopping age $\theta$ uses only that the trajectory is deterministic between events, so it holds for every $g$. Appendix~\ref{app:gaw_verification} in the supplementary material verifies optimality.
\end{proof}
Fig.~\ref{fig:trajectory}(b) traces one delivery cycle. The state idles on $b=0$ until the AoI reaches $z^\star(\Delta)$, and each service run rises with slope one because $\Delta-b$ stays at the busy-start AoI. A run is preempted when it meets the boundary $b=\theta^\star(\Delta-b)$ and delivers if it completes first. Under random arrivals (Fig.~\ref{fig:trajectory}(a)), the idle period and every preemption wait for an arrival, so a run inside the region $\Phi_1\geq 0$ of Theorem~\ref{thm:random_integral_bellman} continues until the next arrival restarts service.
\begin{corollary}[{Reduction to \cite[Thm.~4]{sun2017update}}]\label{cor:no_preemption}
Let preemption be disabled, $g(\Delta)=\Delta$, and $\mathbb{E}[Y^2]<\infty$. Then $v(y)=y\,\mathbb{E}[Y]$ and the idle-phase minimizer is $z^{\star}(\Delta)=\max\{\Delta,\rho-\mathbb{E}[Y]\}$, which is the optimal sampling policy of \cite[Thm.~4]{sun2017update}.
\end{corollary}
\begin{proof}
Every service attempt runs to completion, so \eqref{eq:busy_integral_bellman} admits only $\theta=\infty$ and yields
\begin{equation}\label{acoe:busy_np}
v(y)=Q(y,\infty;\rho)=(y-\rho)\mathbb{E}[Y]+\frac{\mathbb{E}[Y^2]}{2}+\mathbb{E}\!\left[h_I(Y)\right].
\end{equation}
This expression is affine in $y$, so the normalization $v(0)=0$ gives $v(y)=y\,\mathbb{E}[Y]$. Substituting it into \eqref{eq:idle_integral_bellman} yields
\begin{equation}\label{eq28}
h_I(\Delta)=\rho\Delta-\frac{\Delta^2}{2}
+\inf_{z\ge \Delta}\left\{\lambda\kappa_s+\frac{z^2}{2}+(\mathbb{E}[Y]-\rho) z\right\},
\end{equation}
whose minimizer is $z^{\star}(\Delta)=\max\{\Delta,\rho-\mathbb{E}[Y]\}$.
\end{proof}
\enlargethispage{\baselineskip}
The corollary also shows what preemption buys. Without preemption, the second moment $\mathbb{E}[Y^2]$ enters \eqref{acoe:busy_np}, so finite variance is needed. With preemption, any finite threshold $\theta$ caps each service attempt in \eqref{eq:busy_acoe_Q}, so $\mathbb{E}[Y]<\infty$ suffices even when $\mathrm{Var}(Y)=\infty$.

\subsection{Exponential Service Times: Explicit Solutions}
\label{subsec:exp}
Akar \emph{et al.} \cite{akar2026squeezingpreemption} and Arafa \emph{et al.} \cite{arafa2019preemption} optimize constant service-age preemption thresholds for the random-arrival and GaW models, respectively, which leaves open whether an AoI-dependent policy can perform better. We resolve this question for exponentially distributed service times. The following theorem establishes that, under both models, constant service-age thresholds are optimal for any nondecreasing running cost, and that threshold sampling is optimal for linear AoI. 
\begin{theorem}[{Optimality of the Constant Thresholds in \cite{akar2026squeezingpreemption,arafa2019preemption}}]
\label{thm:exp}
Let $Y\sim\text{Exp}(\beta)$, and suppose that the relevant ACOEs of Theorems~\ref{thm:random_integral_bellman} and~\ref{thm:integral_bellman} admit finite Borel solutions of polynomial growth. Define
\begin{equation}
G(x)\triangleq\int_0^\infty e^{-\beta t}\bigl(g(x+t)-\rho\bigr)\,dt,
\end{equation}
with $\rho=\rho_R$ in part (a) and $\rho=\rho_G$ in part (b).
\begin{itemize}
\item[(a)] \emph{Random arrivals:} The busy-phase relative value function separates as
\begin{equation}
h_B(\Delta,b)=G(\Delta)+f(b)
\label{eq:random_yb_separable}
\end{equation}
for a nondecreasing $f$, so the optimal preemption decision depends only on the service age, and an optimal policy is
\begin{equation}
\pi^\star(b)=
\begin{cases}
\textnormal{discard}, & b<b^\star,\\
\textnormal{preempt}, & b\ge b^\star.
\end{cases}
\end{equation}
\item[(b)] \emph{Generate-at-will:} If the infimum in \eqref{eq:busy_integral_bellman} is attained by a measurable selection with $\theta^\star(y)\ge\epsilon$ for all $y$ and some $\epsilon>0$, then $v(y)=G(y)-G(0)$ for $y\ge0$, and a constant preemption threshold $\theta^\star(y)\equiv\theta^\star$ is optimal. If, in addition, $g(\Delta)=\Delta$, the threshold sampling policy $z^\star(\Delta)=\max\{\Delta,\rho-1/\beta\}$ is optimal.
\end{itemize}
Optimality holds under the conditions of Theorems~\ref{thm:random_integral_bellman} and~\ref{thm:integral_bellman}.
\end{theorem}
\begin{proof}
See Appendix~\ref{app:exp_proof} in the supplementary material.
\end{proof}
This simple structure has a clean explanation. The exponential law is memoryless: the residual service time is independent of the service age $b$, so $b$ affects future costs only by setting the AoI upon delivery. The current AoI therefore enters the busy-phase value function only through the additive term $G$, which cancels in the comparison between persisting and preempting. Part~(a) of Theorem~\ref{thm:exp} proves the constant thresholds of \cite{akar2026squeezingpreemption} optimal over all admissible policies, while part~(b) extends the linear-AoI result of \cite[Thm.~2]{aiminpreemption} to general running costs.

\subsection{\textsc{Tailor}: Policy Iteration for General Service Times}
\label{subsec:tailor}
For general $F_Y$, the optimal policies in Theorems~\ref{thm:random_integral_bellman} and~\ref{thm:integral_bellman} admit no closed form. We therefore design \textsc{Tailor}, a policy-iteration algorithm that computes them under both models. 

\textbf{Policy evaluation.} Fix a stationary deterministic policy $\pi\in\Pi_i$ with average cost $\rho^\pi$. Its relative value function, $h_B^\pi$ under random arrivals and $v^\pi$ under GaW, measures the expected cost accumulated from a state in excess of $\rho^\pi$ per unit time and is unique up to an additive constant. Eliminating $h_I$ from the busy-phase ACOE yields a Poisson equation over stages between decision epochs, where $r_\pi(s)$ and $\tau_\pi(s)$ are the expected cost and duration of a stage starting in $s$, and $P_\pi$ averages the relative value over the next starting state.

(a) \emph{Random arrivals:} Substituting \eqref{eq:random_h_idle}, evaluated at the completion state $b+t$, into \eqref{eq:random_h_busy} gives
\begin{equation}\label{eq:poisson_model1}
h_B^\pi(\Delta, b) = r_\pi(\Delta, b) - \rho^\pi \tau_\pi(\Delta, b) + (P_\pi h_B^\pi)(\Delta, b),
\end{equation}
with components \eqref{eq:tau_pi}--\eqref{eq:P_pi} at the bottom of page~\pageref{eq:tau_pi}. Every stage ends at the next Poisson arrival, even if the service completes first, so $\tau_\pi\equiv 1/\nu$.

(b) \emph{Generate-at-will:} For $\pi=(z_\pi,\theta_\pi)$ with waiting target $z_\pi(\Delta)\ge \Delta$, busy-phase threshold $\theta_\pi(y)$, and waiting time $w_\pi(t)\triangleq z_\pi(t)-t\ge 0$, the idle-phase ACOE reduces to
\begin{equation}\label{eq:hI_pi}
h_I^\pi(t)=\lambda\kappa_s+v^\pi(z_\pi(t))+\int_{t}^{t+w_\pi(t)} g(s) \, ds-\rho^\pi w_\pi(t),
\end{equation}
where the integral equals $t\,w_\pi(t)+\tfrac12 w_\pi(t)^2$ for $g(\Delta)=\Delta$. Inserting \eqref{eq:hI_pi} into the busy-phase ACOE $v^\pi(y)=Q(y,\theta_\pi(y);\rho^\pi)$ gives
\begin{equation}\label{eq:poisson}
v^\pi(y)=r_\pi(y)-\rho^\pi \tau_\pi(y) + (P_\pi v^\pi)(y),\qquad v^\pi(0)=0,
\end{equation}
with components given in \eqref{eq:tau_pi_gaw}--\eqref{eq:P_pi_gaw} at the bottom of page~\pageref{eq:tau_pi_gaw}, where the $\bar F_Y(\theta_\pi(y))$ terms vanish if $\theta_\pi(y)=\infty$.

\begin{figure*}[!t]
\normalsize
\hrulefill
\vspace*{-2pt}
\begin{subequations}\label{eq:bellman_components_model1}
\begin{align}
\tau_\pi(\Delta,b) &= \int_0^\infty e^{-\nu t}S_b(t)\,dt+\frac{1}{\nu}\int_{[0,\infty)}e^{-\nu t}\,dF_{R_b}(t)=\frac{1}{\nu},\label{eq:tau_pi}\\
r_\pi(\Delta,b) &= \int_0^\infty e^{-\nu t}S_b(t)\,g(\Delta+t)\,dt+\lambda\kappa_p\int_0^\infty \nu e^{-\nu t}S_b(t)\,\mathbf 1_{\{\pi(\Delta+t,b+t)=\preemptaction\}}\,dt\notag\\
&\quad+\int_{[0,\infty)}e^{-\nu t}\Big(\int_0^\infty e^{-\nu s}g(b+t+s)\,ds+\lambda\kappa_s\Big)\,dF_{R_b}(t),\label{eq:r_pi}\\
(P_\pi h_B^\pi)(\Delta,b) &= \int_{[0,\infty)}\int_0^\infty \nu e^{-\nu(t+s)}\,h_B^\pi(b+t+s,0)\,ds\,dF_{R_b}(t)\notag\\
&\quad+\int_0^\infty \nu e^{-\nu t}S_b(t)\Big(\mathbf 1_{\{\pi(\Delta+t,b+t)=\textsf{discard}\}}\,h_B^\pi(\Delta+t,b+t)+\mathbf 1_{\{\pi(\Delta+t,b+t)=\preemptaction\}}\,h_B^\pi(\Delta+t,0)\Big)\,dt.\label{eq:P_pi}
\end{align}
\end{subequations}
\par\vspace*{6pt}\hrule height 0.2pt\vspace*{-4pt}
\begin{subequations}\label{eq:bellman_components_gaw}
\begin{align}
\tau_\pi(y) &= A\big(\theta_\pi(y)\big)+\int_{[0,\theta_\pi(y)]} w_\pi(t)\,dF_Y(t),\label{eq:tau_pi_gaw}\\
r_\pi(y) &= \int_0^{\theta_\pi(y)} g(y+t)\,\bar F_Y(t)\,dt+\lambda\kappa_p\,\bar F_Y\big(\theta_\pi(y)\big)+\int_{[0,\theta_\pi(y)]}\Big(\lambda\kappa_s+\int_t^{z_\pi(t)} g(s)\,ds\Big)\,dF_Y(t),\label{eq:r_pi_gaw}\\
(P_\pi v^\pi)(y) &= \int_{[0,\theta_\pi(y)]} v^\pi\big(z_\pi(t)\big)\,dF_Y(t)+\bar F_Y\big(\theta_\pi(y)\big)\,v^\pi\big(y+\theta_\pi(y)\big).\label{eq:P_pi_gaw}
\end{align}
\end{subequations}
\end{figure*}

\textbf{Policy improvement.} Given $(\rho^{\pi_k},h^{\pi_k})$, \textsc{Tailor} applies the preemption rule \eqref{eq:random_preempt_rule} under random arrivals, or the sampling and preemption rules \eqref{eq:zstar_def} and \eqref{eq:taustar_def} under GaW, with $(\rho,h)$ replaced by $(\rho^{\pi_k},h^{\pi_k})$. It keeps the current action whenever that action attains the minimum and stops once the policy no longer changes; on the finite model below, this rule guarantees finite termination at a policy that attains $\rho_i(g,\lambda)$~\cite[Ch.~11]{puterman1994markov}.

\textbf{Truncation and hybrid action grid.} We truncate the unbounded state space to solve \eqref{eq:poisson_model1} and \eqref{eq:poisson} under heavy-tailed delays. Under GaW, we discretize the busy-start AoI $y$ and waiting target $z$ on $[0,y_{\mathrm{cut}}]$ with step $dt$, and choose preemption ages from a hybrid grid $\Theta_h\subset(0,y_{\mathrm{cut}}]$ comprising all positive $dt$-grid points up to $\theta_{\mathrm{fine}}$ and log-spaced points rounded to that grid thereafter, limiting the action set under heavy tails. Under random arrivals, actions remain binary, and we discretize the triangle $0\le b\le\Delta\le\Delta_{\mathrm{cut}}$ with step $dt$. We round service completions and arrivals up to step ends, with completions processed first and arrival probability $1-e^{-\nu dt}$ per step. We integrate $g$ exactly along unit-rate AoI trajectories within each step. Together with the far-field rule below, these grids and transitions define a semi-Markov decision process (SMDP)~\cite{puterman1994markov} with finitely many states and actions, which we call the finite model; \textsc{Tailor} solves it, and the computational guarantees of Section~\ref{sec:tail} refer to it. Transitions beyond the cutoff still need relative values off the grid.

\textbf{Executable far-field closure.} We obtain the required off-grid values from the exact relative continuation cost of an executable policy that restarts unfinished service until delivery. For the two policies below, bounded or exponential restart clocks yield a geometric number of attempts, so the delivery time $D$ has finite moments of all orders even when $\mathbb{E}[Y^2]=\infty$. Under GaW, the closure policy restarts each unfinished attempt at a fixed age $\theta_0\in(0,y_{\mathrm{cut}}]$ with $p_0\triangleq F_Y(\theta_0)>0$. From busy-start AoI $y\ge y_{\mathrm{cut}}$, its relative cost is
\begin{equation}\label{eq:tail_closure}
\begin{aligned}
v_{\mathrm{tail}}(y)={}&\sum_{k=0}^{\infty}(1-p_0)^k\int_0^{\theta_0}\bar F_Y(s)\big(g(y+k\theta_0+s)-\rho\big)\,ds\\
&+\lambda\kappa_p\frac{1-p_0}{p_0}+\frac{1}{p_0}\int_{[0,\theta_0]}h_I(t)\,dF_Y(t),
\end{aligned}
\end{equation}
whose three terms are the running cost in excess of $\rho$, the expected preemption penalty, and the expected idle-phase value upon delivery, computed from in-grid values of $h_I$ in the finite model. The series converges for polynomially growing $g$. For fixed $\theta_0$, \eqref{eq:tail_closure} is affine in $(\rho,h_I)$, so substituting $v_{\mathrm{tail}}(y+\theta)$ for $v(y+\theta)$ beyond $y_{\mathrm{cut}}$ keeps the evaluation system linear. Under random arrivals, the closure policy instead preempts at every busy arrival until delivery, so each attempt succeeds with probability $p_\nu\triangleq\mathbb{E}[e^{-\nu Y}]$. From busy-start AoI $x>\Delta_{\mathrm{cut}}$, the corresponding relative cost is
\begin{equation}\label{eq:tail_closure_random}
h_{\mathrm{tail}}(x)=\mathbb{E}\Big[\int_0^{D}\big(g(x+t)-\rho\big)\,dt\Big]+\lambda\kappa_p\frac{1-p_\nu}{p_\nu}+c,
\end{equation}
where $c\triangleq\mathbb{E}[h_I(W)]$ and $W$ is the service time of the update delivered under this policy. Since the next arrival can start service beyond $\Delta_{\mathrm{cut}}$, $c$ couples back to $h_{\mathrm{tail}}$ through a scalar linear equation derived in Appendix~\ref{app:far_field_random} in the supplementary material, which we append to the evaluation system. More generally, the far-field rule may preempt at busy arrivals only when the service age is at least $k\in[0,\infty]$, covering Always-Preempt ($k=0$) and Never-Preempt ($k=\infty$).

The closures \eqref{eq:tail_closure} and \eqref{eq:tail_closure_random} adapt to $g$. For $g(\Delta)=\Delta$, $v_{\mathrm{tail}}$ and $h_{\mathrm{tail}}$ are affine in the busy-start AoI with slope $\mathbb{E}[D]$, equal to $A(\theta_0)/p_0$ and $(1-p_\nu)/(\nu p_\nu)$, respectively. Formally, letting $\theta_0\to\infty$ in \eqref{eq:tail_closure} recovers the never-preempt comparison cost in \cite[Lemma~5]{aiminpreemption} when $\mathbb{E}[Y^2]<\infty$ and $\mathbb{E}[|h_I(Y)|]<\infty$. For the VaR indicator cost, both closures are constant above $\Delta_{\mathrm{th}}$, and for the CVaR hinge cost, they are affine with slope $\mathbb{E}[D]$ above $\eta$. The far-field closures supply boundary values, while \textsc{Tailor} improves actions only on the grid. We enlarge the cutoff until the computed average cost stabilizes. Algorithm~\ref{alg:tailor} summarizes the procedure, which extends \cite[Algorithm~1]{aiminpreemption} to random arrivals with these executable far-field closures.

\begin{algorithm}[t]
\caption{\textsc{Tailor}: Policy Iteration for Problem~\ref{p1}}
\label{alg:tailor}
\KwIn{Model $i\in\{R,G\}$, service distribution $F_Y$, running cost $g$, weight $\lambda$, grid parameters $(dt,y_{\mathrm{cut}},\Delta_{\mathrm{cut}},\theta_{\mathrm{fine}})$, restart age $\theta_0$.}
\KwOut{Average cost $\rho_i(g,\lambda)$ and an optimal stationary policy $\pi^\star$.}
Build the truncated state and hybrid action grids\;
Initialize a stationary policy $\pi_0\in\Pi_i$\;
\For{$k=0,1,2,\dots$}{
\textbf{Evaluation:} solve \eqref{eq:poisson_model1} closed by \eqref{eq:tail_closure_random} if $i=R$, or \eqref{eq:poisson} closed by \eqref{eq:tail_closure} if $i=G$, for $(\rho^{\pi_k},h^{\pi_k})$\;
\textbf{Improvement:} obtain $\pi_{k+1}$ from \eqref{eq:random_preempt_rule} if $i=R$, or from \eqref{eq:zstar_def} and \eqref{eq:taustar_def} if $i=G$, keeping the current action under ties\;
\lIf{$\pi_{k+1}=\pi_k$}{\Return $(\rho^{\pi_k},\pi_k)$}
}
\end{algorithm}

\section{Budget-Constrained Tail-Risk Freshness Optimization}
\label{sec:tail}
Minimizing the average AoI does not prevent occasional severe excursions, because a low long-run mean dilutes rare episodes of extreme staleness. In industrial automation and teleoperation, these episodes rather than the average govern system safety: once the AoI crosses a safety threshold, the controller acts on obsolete state feedback, which may destabilize the loop. Such applications must therefore bound the age exceeded during the worst $1-\alpha$ fraction of time and the average age during those excursions, which are exactly the VaR and the CVaR of the AoI. This section minimizes either tail metric under an action-cost budget, using the same sampling and preemption actions.

\subsection{Problem Formulation}
\label{subsec:risk_formulations}
For $\pi\in\Pi_i$, $i\in\{R,G\}$, let
\begin{equation}
\mu_\pi(\mathcal B)\triangleq\lim_{T\to\infty}\frac{1}{T}\,\mathbb E^\pi\!\left[\int_0^T\mathbf 1_{\{\Delta(t)\in\mathcal B\}}\,dt\right]
\end{equation}
be the long-run AoI occupation measure, and let $\tilde\Delta_\pi\sim\mu_\pi$. At confidence level $\alpha\in(0,1)$, define
\begin{align}
\operatorname{VaR}_\alpha(\tilde{\Delta}_\pi) &\triangleq \inf\big\{x\ge 0:\mathbb P(\tilde\Delta_\pi>x)\le 1-\alpha\big\},\label{eq:var_def}\\
\operatorname{CVaR}_\alpha(\tilde{\Delta}_\pi) &\triangleq \mathbb E\big[\tilde\Delta_\pi\bigm|\tilde\Delta_\pi\ge\operatorname{VaR}_\alpha(\tilde\Delta_\pi)\big].\label{eq:cvar_def}
\end{align}
As illustrated in Fig.~\ref{fig:VaR_CVaR_demo}, VaR identifies the age level exceeded during the worst $1-\alpha$ fraction of time, whereas CVaR measures the average age during those excursions.

\begin{problem}[Constrained VaR Minimization]\label{p2}
\begin{equation}
V^\star_{\bar K_{\max},\alpha,i}=\inf_{\pi\in\Pi_i}\ \operatorname{VaR}_\alpha(\tilde\Delta_\pi)\quad\textnormal{s.t.}\quad \bar K_\pi\le\bar K_{\max}.
\end{equation}
\end{problem}

\begin{problem}[Constrained CVaR Minimization]\label{p3}
\begin{equation}
C^\star_{\bar K_{\max},\alpha,i}=\inf_{\pi\in\Pi_i}\ \operatorname{CVaR}_\alpha(\tilde\Delta_\pi)\quad\textnormal{s.t.}\quad \bar K_\pi\le\bar K_{\max}.
\end{equation}
\end{problem}

\begin{remark}[Difficulty Beyond the Mean]
The ACOEs of Section~\ref{sec:age_optimal_preemption} exploit the fact that the mean AoI accumulates stage by stage. The VaR and the CVaR are a quantile and a tail average of the stationary AoI distribution, which shift whenever the policy changes, so neither splits into per-stage costs and the ACOEs do not apply directly. Augmenting the state with risk variables, as Zhou \emph{et al.}~\cite{zhou2020risk} do for the CVaR of discounted AoI, restores additivity, but every added dimension multiplies the grid that \textsc{Tailor} must solve, which becomes impractical at the resolution that tail metrics require.

\end{remark}

\subsection{Thresholding the Tail into Average Costs}
\label{subsec:reduction}
We avoid state augmentation by expressing each tail metric as an average running cost through a scalar threshold. This leaves the action-cost budget, which we remove through Lagrangian duality (Section~\ref{subsec:kernel}). At every candidate threshold, both tail problems therefore reduce to the following budget-constrained average-cost problem.
\begin{problem}[Budget-Constrained Average Cost]\label{p4}
For a running cost $g$,
\begin{equation}\label{eq:budget_value}
J_i(g;\bar K_{\max})\triangleq\inf_{\pi\in\Pi_i}\ \bar g_\pi\quad\textnormal{s.t.}\quad \bar K_\pi\le\bar K_{\max},
\end{equation}
where $\bar g_\pi$ is the long-run average of $g(\Delta(t))$ and $\bar K_\pi$ is the long-run average action cost, as in Problem~\ref{p1}.
\end{problem}
The two tail metrics differ in how the threshold enters $g$.

\subsubsection{Transforming VaR into a Threshold-Crossing Problem}
A quantile of the AoI distribution is a fraction of time in disguise. By \eqref{eq:var_def}, a policy has VaR at most $\Delta_{\mathrm{th}}$ if and only if $\mathbb P(\tilde\Delta_\pi>\Delta_{\mathrm{th}})\le1-\alpha$, and the occupation measure makes it a time average:
\begin{equation}\label{eq:var_avg}
\mathbb P(\tilde\Delta_\pi>\Delta_{\mathrm{th}})=\lim_{T\to\infty}\frac{1}{T}\,\mathbb E^\pi\!\left[\int_0^T\mathbf 1_{\{\Delta(t)>\Delta_{\mathrm{th}}\}}\,dt\right].
\end{equation}
The right-hand side is the average running cost $\bar g_\pi$ of Problem~\ref{p1} with $g(\Delta)=\mathbf 1_{\{\Delta>\Delta_{\mathrm{th}}\}}$, the fraction of time that the AoI spends above $\Delta_{\mathrm{th}}$. We thereby transform Problem~\ref{p2} into finding the smallest threshold whose minimum violation fraction under the budget is at most $1-\alpha$:
\begin{equation}\label{eq:var_outer}
V^\star_{\bar K_{\max},\alpha,i}=\inf\{\Delta_{\mathrm{th}}\colon J_i(\mathbf 1_{\{\Delta>\Delta_{\mathrm{th}}\}};\bar K_{\max})\le1-\alpha\}.
\end{equation}
This equality requires the infimum in $J_i$ to be attained, which Lemma~\ref{prop:kernel} ensures on the finite model of Section~\ref{subsec:tailor}. Testing each candidate threshold thus requires one instance of Problem~\ref{p4}, which evaluates $J_i$ in \eqref{eq:budget_value} with $g(\Delta)=\mathbf 1_{\{\Delta>\Delta_{\mathrm{th}}\}}$.

\subsubsection{Transforming CVaR into a Scalar Minimization}
A tail average is an ordinary average cost minimized over a single scalar. We use the \emph{Rockafellar--Uryasev representation}~\cite[Thm.~1]{rockafellar2000cvar}, which writes the CVaR as a minimization over a scalar threshold:
\begin{equation}\label{eq:cvar_ru}
\operatorname{CVaR}_\alpha(\tilde{\Delta}_\pi)=\min_{\eta\ge0}\Big\{\underbrace{\eta+\frac{\mathbb E\big[(\tilde\Delta_\pi-\eta)^+\big]}{1-\alpha}}_{\triangleq F_\pi(\eta)}\Big\},
\end{equation}
where $(u)^+\triangleq\max\{u,0\}$ and the minimum is attained at $\eta=\operatorname{VaR}_\alpha(\tilde\Delta_\pi)$: since $F_\pi'(\eta)=1-\mathbb P(\tilde\Delta_\pi>\eta)/(1-\alpha)$, the slope vanishes exactly where the AoI exceeds $\eta$ for a $1-\alpha$ fraction of time. The occupation measure turns the expectation into a time average:
\begin{equation}\label{eq:cvar_avg}
\mathbb E\big[(\tilde\Delta_\pi-\eta)^+\big]=\lim_{T\to\infty}\frac{1}{T}\,\mathbb E^\pi\!\left[\int_0^T(\Delta(t)-\eta)^+\,dt\right].
\end{equation}
The right-hand side is the average running cost $\bar g_\pi$ of Problem~\ref{p1} with the hinge cost $g(\Delta)=(\Delta-\eta)^+$, which charges only the ages above $\eta$ and weighs how far they exceed it. We substitute \eqref{eq:cvar_ru} into Problem~\ref{p3} and exchange the two infima, which transforms Problem~\ref{p3} into minimizing a function of the single scalar $\eta$:
\begin{align}
C^\star_{\bar K_{\max},\alpha,i}&\overset{\eqref{eq:cvar_ru}}{=}\inf_{\substack{\pi\in\Pi_i\\\bar K_\pi\le\bar K_{\max}}}\;\min_{\eta\ge0}F_\pi(\eta)\nonumber\\
&=\inf_{\eta\ge0}\;\inf_{\substack{\pi\in\Pi_i\\\bar K_\pi\le\bar K_{\max}}}\;F_\pi(\eta)\nonumber\\
&\overset{\eqref{eq:budget_value},\eqref{eq:cvar_avg}}{=}\inf_{\eta\ge0}\Big\{\underbrace{\eta+\frac{J_i\big((\Delta-\eta)^+;\bar K_{\max}\big)}{1-\alpha}}_{\triangleq H_i(\eta)}\Big\}.\label{eq:cvar_outer}
\end{align}
For each candidate $\eta$, one instance of Problem~\ref{p4} gives the smallest hinge cost that any budget-feasible policy attains, so $H_i(\eta)$ is the smallest $F_\pi(\eta)$ among these policies. Unlike the VaR, which reduces to locating where a nonincreasing curve crosses a level, the CVaR requires minimizing $H_i$ itself.

\subsection{Lagrangian Duality}
\label{subsec:kernel}
Both \eqref{eq:var_outer} and \eqref{eq:cvar_outer} require solving Problem~\ref{p4} at every candidate threshold. We remove its budget constraint with a Lagrange multiplier and show that no optimality is lost (Lemma~\ref{prop:duality}). For a fixed multiplier, evaluating the dual function reduces to an unconstrained average-cost problem, an instance of Problem~\ref{p1} that \textsc{Tailor} solves directly. 

\subsubsection{Formulating the Lagrangian Dual of Problem~\ref{p4}}
We attach a multiplier $\lambda\ge0$ to the budget constraint. The Lagrangian $\bar g_\pi+\lambda(\bar K_\pi-\bar K_{\max})$ never exceeds $\bar g_\pi$ when $\pi$ meets the budget, so its infimum over all policies bounds $J_i(g;\bar K_{\max})$ from below for every $\lambda$. The best bound defines a new problem.
\begin{problem}[Lagrangian Dual of Problem~\ref{p4}]\label{p5}
For a running cost $g$,
\begin{align}
D_i(g;\bar K_{\max})&\triangleq\sup_{\lambda\ge0}\ \underbrace{\inf_{\pi\in\Pi_i}\big\{\bar g_\pi+\lambda(\bar K_\pi-\bar K_{\max})\big\}}_{\triangleq\, d_i(g,\lambda;\bar K_{\max})}\nonumber\\
&=\sup_{\lambda\ge0}\ \Big\{\inf_{\pi\in\Pi_i}\big\{\bar g_\pi+\lambda\bar K_\pi\big\}-\lambda\bar K_{\max}\Big\}\nonumber\\
&=\sup_{\lambda\ge0}\ \big\{\rho_i(g,\lambda)-\lambda\bar K_{\max}\big\},\label{eq:dual_function}
\end{align}
where $d_i(g,\lambda;\bar K_{\max})$ is the dual function.
\end{problem}
Here, $\rho_i(g,\lambda)$ is the optimal value of Problem~\ref{p1}, and the first equality in \eqref{eq:dual_function} holds because $\lambda\bar K_{\max}$ does not depend on $\pi$, so each evaluation of $d_i(g,\lambda;\bar K_{\max})$ is one \textsc{Tailor} solve. Every $\lambda\ge0$ thus gives $d_i(g,\lambda;\bar K_{\max})\le J_i(g;\bar K_{\max})$, which is \textit{weak duality}, and $d_i(g,\lambda;\bar K_{\max})$ is concave in $\lambda$ as an infimum of affine functions. We next show that the duality gap vanishes, which transforms Problem~\ref{p4} into Problem~\ref{p5}, a concave maximization over the single scalar $\lambda$.
\begin{lemma}[Strong Duality]
\label{prop:duality}
Suppose that some $\pi_s\in\Pi_i$ satisfies $\bar K_{\pi_s}<\bar K_{\max}$ and $\bar g_{\pi_s}<\infty$. Then
\begin{equation}\label{eq:strong_duality}
J_i(g;\bar K_{\max})=D_i(g;\bar K_{\max})=\max_{\lambda\ge0}d_i(g,\lambda;\bar K_{\max}),
\end{equation}
\end{lemma}
\begin{proof}
See Appendix~\ref{app:duality_proof} in the supplementary material.
\end{proof}
Under GaW, the condition of Lemma~\ref{prop:duality} holds for any $\bar K_{\max}>0$. Under random arrivals, the policy that never preempts incurs an average action cost of $\kappa_s\nu/(1+\nu\mathbb E[Y])$. Thus, the condition holds whenever $\bar K_{\max}$ exceeds this cost (Lemma~\ref{lem:slater} in the supplementary material).

\subsubsection{Solving Problem~\ref{p5} by Multiplier Bisection}
From here on, we work on the finite model of Section~\ref{subsec:tailor}, which discretizes the AoI and the service age with step $dt$ and closes the grid with the far-field rule, so that it has finitely many states and actions. On this model, for every $\lambda\ge0$, \textsc{Tailor} terminates in finitely many steps at a deterministic stationary policy $\pi_\lambda$ that attains $\rho_i(g,\lambda)$ (Lemma~\ref{prop:kernel} in the supplementary material). Let $\bar K_i(g,\lambda)\triangleq\bar K_{\pi_\lambda}$ denote the average action cost of $\pi_\lambda$, and let
\begin{equation}\label{eq:lambda_star}
\lambda^\star\triangleq\inf\{\lambda\ge0\colon\bar K_i(g,\lambda)\le\bar K_{\max}\}.
\end{equation}
The next lemma lets us locate a dual maximizer by bisection.
\begin{lemma}[Dual Structure and Threshold Multiplier]
\label{lem:spending}
The following assertions hold true:
\begin{itemize}
\item[(i)] $\bar K_i(g,\lambda)$ is nonincreasing and piecewise constant in $\lambda$.
\item[(ii)] $d_i(g,\cdot\,;\bar K_{\max})$ is concave and piecewise linear. It is nondecreasing on $[0,\lambda]$ if $\bar K_i(g,\lambda)>\bar K_{\max}$ and nonincreasing on $[\lambda,\infty)$ if $\bar K_i(g,\lambda)<\bar K_{\max}$.
\item[(iii)] If $\lambda^\star<\infty$, then $\lambda^\star$ maximizes
$d_i(g,\cdot\,;\bar K_{\max})$, regardless of whether
$\bar K_i(g,\lambda^\star)\le\bar K_{\max}$.
\item[(iv)] If $g\ge0$ and there exists a policy $\pi_s$ such that
$\bar K_{\pi_s}<\bar K_{\max}$ and $\bar g_{\pi_s}<\infty$, then
$\lambda^\star<\infty$.
\end{itemize}
\end{lemma}
\begin{proof}
See Appendix~\ref{app:spending_proof} in the supplementary material.
\end{proof}
Lemma~\ref{lem:spending} reduces Problem~\ref{p5} to locating the threshold $\lambda^\star$ in \eqref{eq:lambda_star}. By part~(i), the optimal action cost $\bar K_i(g,\lambda)$ is nonincreasing in $\lambda$. By part~(ii), the dual function is nondecreasing while this cost exceeds the budget and nonincreasing once it falls below, so $\lambda^\star$ maximizes the dual function by part~(iii), and part~(iv) ensures that $\lambda^\star<\infty$.

We locate $\lambda^\star$ by bisection. If $\pi_0$ meets the budget, it already solves Problem~\ref{p4}. Otherwise, starting from $\lambda_{\mathrm{lo}}=0$ and $\lambda_{\mathrm{hi}}=1$, we double the upper multiplier, moving each unsuccessful multiplier to $\lambda_{\mathrm{lo}}$, until its optimal policy meets the budget, which takes finitely many steps by part~(iv). We then bisect $[\lambda_{\mathrm{lo}},\lambda_{\mathrm{hi}}]$ with one \textsc{Tailor} solve per midpoint, keeping $\bar K_{\pi_{\mathrm{lo}}}>\bar K_{\max}\ge\bar K_{\pi_{\mathrm{hi}}}$ for $\pi_{\mathrm{lo}}\triangleq\pi_{\lambda_{\mathrm{lo}}}$ and $\pi_{\mathrm{hi}}\triangleq\pi_{\lambda_{\mathrm{hi}}}$.

Because $\bar K_i(g,\cdot)$ is piecewise constant by part~(i) of Lemma~\ref{lem:spending}, it can jump across the budget at $\lambda^\star$, so neither endpoint policy need exhaust the budget. We therefore use the \emph{random mixture}~\cite{beutler1985optimal,altman1999constrained} $\hat\pi$ that selects $\pi_{\mathrm{lo}}$ with probability $q$ at time zero and $\pi_{\mathrm{hi}}$ otherwise, mixing their long-run averages linearly. Choosing
\begin{equation}\label{eq:budget_mixture}
q\triangleq\frac{\bar K_{\max}-\bar K_{\pi_{\mathrm{hi}}}}{\bar K_{\pi_{\mathrm{lo}}}-\bar K_{\pi_{\mathrm{hi}}}},\qquad\mu_{\hat\pi}=q\mu_{\pi_{\mathrm{lo}}}+(1-q)\mu_{\pi_{\mathrm{hi}}},
\end{equation}
gives $\hat\pi$ an average action cost of exactly $\bar K_{\max}$. Since $\hat\pi$ meets the budget, its average running cost is an upper bound on $J_i(g;\bar K_{\max})$, and by weak duality every visited dual value is a lower bound. This gap at least halves with each bisection step (Lemma~\ref{lem:bisection_gap} in the supplementary material). Algorithm~\ref{alg:kernel} summarizes the procedure. It returns $\hat\pi$ with the bracket $\underline J(g)\le J_i(g;\bar K_{\max})\le\overline J(g)$, where $\overline J(g)\triangleq\bar g_{\hat\pi}$ and $\underline J(g)$ is the largest visited dual value, and it reaches any bracket width $\delta>0$ within $O(\log(1/\delta))$ solves of \textsc{Tailor}.

\begin{algorithm}[t]
\caption{Multiplier Bisection for Problem~\ref{p4}}
\label{alg:kernel}
\KwIn{Model $i$, running cost $g$, budget $\bar K_{\max}$, and tolerance $\delta$.}
\KwOut{A budget-feasible policy $\hat\pi$ and a bracket $[\underline J(g),\overline J(g)]$ on $J_i(g;\bar K_{\max})$.}
Solve \textsc{Tailor} at $\lambda=0$\;
\lIf{$\bar K_{\pi_0}\le\bar K_{\max}$}{\Return $\pi_0$ and $[\bar g_{\pi_0},\bar g_{\pi_0}]$}
$\lambda_{\mathrm{lo}}\leftarrow0$, $\lambda_{\mathrm{hi}}\leftarrow1$, and solve \textsc{Tailor} at $\lambda_{\mathrm{hi}}$\;
\lWhile{$\bar K_{\pi_{\lambda_{\mathrm{hi}}}}>\bar K_{\max}$}{$\lambda_{\mathrm{lo}}\leftarrow\lambda_{\mathrm{hi}}$, $\lambda_{\mathrm{hi}}\leftarrow2\lambda_{\mathrm{hi}}$, and solve \textsc{Tailor} at $\lambda_{\mathrm{hi}}$}
Form $\hat\pi$ by \eqref{eq:budget_mixture}, $\overline J\leftarrow\bar g_{\hat\pi}$, and $\underline J\leftarrow$ the largest visited dual value\;
\While{$\overline J-\underline J>\delta$}{
Solve \textsc{Tailor} at $\lambda_{\mathrm m}\leftarrow(\lambda_{\mathrm{lo}}+\lambda_{\mathrm{hi}})/2$\;
\lIf{$\bar K_{\pi_{\lambda_{\mathrm m}}}>\bar K_{\max}$}{$\lambda_{\mathrm{lo}}\leftarrow\lambda_{\mathrm m}$}\lElse{$\lambda_{\mathrm{hi}}\leftarrow\lambda_{\mathrm m}$}
Form $\hat\pi$ by \eqref{eq:budget_mixture} and update $\overline J$ and $\underline J$\;
}
\Return $\hat\pi$ and $[\underline J,\overline J]$\;
\end{algorithm}

\subsection{VaR Minimization via Threshold Bisection}
\label{subsec:var}
We test a candidate threshold $\Delta_{\mathrm{th}}$ by checking whether the minimum violation probability $J_i(\mathbf 1_{\{\Delta>\Delta_{\mathrm{th}}\}};\bar K_{\max})$ in \eqref{eq:var_outer} is at most $1-\alpha$. This probability is nonincreasing in $\Delta_{\mathrm{th}}$, so the thresholds that pass the test form an upper interval whose lower endpoint, the optimal VaR, we locate by bisection.

\subsubsection{Threshold Test}
At each threshold, Algorithm~\ref{alg:kernel} returns a budget-feasible policy $\hat\pi$ and bounds $\underline J\le J_i(\mathbf 1_{\{\Delta>\Delta_{\mathrm{th}}\}};\bar K_{\max})\le\overline J$, where, for this indicator cost, $\overline J=\bar g_{\hat\pi}$ is the violation probability of $\hat\pi$ itself. If $\overline J\le1-\alpha$, the policy $\hat\pi$ keeps the AoI above $\Delta_{\mathrm{th}}$ for at most a $1-\alpha$ fraction of time, so its VaR is at most $\Delta_{\mathrm{th}}$ and the threshold passes. If $\underline J>1-\alpha$, no budget-feasible policy keeps the violation probability within $1-\alpha$, so the threshold fails and, by monotonicity, $V^\star_{\bar K_{\max},\alpha,i}\ge\Delta_{\mathrm{th}}$. Otherwise, $\underline J\le1-\alpha<\overline J$, and Algorithm~\ref{alg:kernel} keeps bisecting the multiplier until one test resolves or $\overline J-\underline J\le\delta$, in which case the search stops without moving either endpoint and returns its current bracket on $V^\star_{\bar K_{\max},\alpha,i}$.

\subsubsection{Initial Bracket}
We start the bisection from a lower bound on the optimal VaR. Between consecutive deliveries, the AoI grows at unit rate and spends at most $b-a$ time in any interval $(a,b]$, so the delivery rate limits the long-run time the AoI spends there. The delivery rate is at most $\bar r_R\triangleq\nu$ under random arrivals and $\bar r_G\triangleq\bar K_{\max}/\kappa_s$ under GaW, where $\kappa_s>0$ is the cost per sample. Recall that $\mu_\pi$ is the long-run fraction of time the AoI spends in each age set under $\pi$, and $\tilde\Delta_\pi\sim\mu_\pi$.
\begin{lemma}[Delivery-Rate Limit on AoI Concentration]
\label{prop:density}
Every budget-feasible $\pi\in\Pi_i$ satisfies
\begin{equation}\label{eq:search_density_bound}
\mu_\pi((a,b])\le\bar r_i(b-a),\qquad 0\le a<b,
\end{equation}
so $\mu_\pi$ has a density bounded by $\bar r_i$ and $\operatorname{VaR}_\alpha(\tilde\Delta_\pi)\ge\alpha/\bar r_i$.
\end{lemma}
\begin{proof}
See Appendix~\ref{app:density_proof} in the supplementary material.
\end{proof}
Lemma~\ref{prop:density} gives $\Delta_{\mathrm{lo}}=\alpha/\bar r_i\le V^\star_{\bar K_{\max},\alpha,i}$, and this lower bound does not depend on the service-time distribution. Any budget-feasible reference policy $\pi_{\mathrm{ref}}$ given to Algorithm~\ref{alg:var_search} yields $\Delta_{\mathrm{hi}}=\operatorname{VaR}_\alpha(\tilde\Delta_{\pi_{\mathrm{ref}}})\ge V^\star_{\bar K_{\max},\alpha,i}$.

\subsubsection{Bisection}
\textsc{Bisec-Tailor} (Algorithm~\ref{alg:var_search}) tests the midpoint of $[\Delta_{\mathrm{lo}},\Delta_{\mathrm{hi}}]$. A pass moves $\Delta_{\mathrm{hi}}$ to the midpoint and saves the returned policy as $\pi_{\mathrm{save}}$, and a failure moves $\Delta_{\mathrm{lo}}$ there, so $\Delta_{\mathrm{lo}}\le V^\star_{\bar K_{\max},\alpha,i}\le\Delta_{\mathrm{hi}}$ holds throughout. On the finite model, $\pi_{\mathrm{save}}$ is budget-feasible with VaR at most $\Delta_{\mathrm{hi}}$, so its VaR exceeds the optimum by at most $\Delta_{\mathrm{hi}}-\Delta_{\mathrm{lo}}$, and each resolved test halves the bracket width and hence this bound. After $n$ resolved tests, $\pi_{\mathrm{save}}$ is therefore within $2^{-n}\big(\operatorname{VaR}_\alpha(\tilde\Delta_{\pi_{\mathrm{ref}}})-\alpha/\bar r_i\big)$ of the optimal VaR.

\begin{algorithm}[t]
\caption{\textsc{Bisec-Tailor} for VaR}
\label{alg:var_search}
\KwIn{Model $i$, $\alpha$, $\bar K_{\max}$, tolerances $\epsilon$ and $\delta$, and a budget-feasible policy $\pi_{\mathrm{ref}}$.}
\KwOut{A budget-feasible policy and a bracket on $V^\star_{\bar K_{\max},\alpha,i}$ that bounds its gap.}
$[\Delta_{\mathrm{lo}},\Delta_{\mathrm{hi}}]\leftarrow[\alpha/\bar r_i,\operatorname{VaR}_\alpha(\tilde\Delta_{\pi_{\mathrm{ref}}})]$; $\pi_{\mathrm{save}}\leftarrow\pi_{\mathrm{ref}}$\;
\While{$\Delta_{\mathrm{hi}}-\Delta_{\mathrm{lo}}>\epsilon$}{
$\Delta_{\mathrm{th}}\leftarrow(\Delta_{\mathrm{lo}}+\Delta_{\mathrm{hi}})/2$; run Algorithm~\ref{alg:kernel} on $\mathbf 1_{\{\Delta>\Delta_{\mathrm{th}}\}}$ until a test resolves or $\overline J-\underline J\le\delta$\;
\lIf{$\overline J\le1-\alpha$}{$\Delta_{\mathrm{hi}}\leftarrow\Delta_{\mathrm{th}}$ and $\pi_{\mathrm{save}}\leftarrow\hat\pi$}
\lElseIf{$\underline J>1-\alpha$}{$\Delta_{\mathrm{lo}}\leftarrow\Delta_{\mathrm{th}}$}
\lElse{\textbf{break}}
}
\Return $\pi_{\mathrm{save}}$ and $[\Delta_{\mathrm{lo}},\Delta_{\mathrm{hi}}]$\;
\end{algorithm}

\subsection{CVaR Minimization via Global Interval Search}
\label{subsec:cvar}
Minimizing $H_i$ in \eqref{eq:cvar_outer} requires a global search: $H_i$ is the lower envelope of the convex functions $F_\pi$ over budget-feasible policies, and since these curves cross, $H_i$ may have several local minima. A local search may stop at a suboptimal one, and a grid cannot exclude lower values between its points. We design \textsc{Interval-Tailor}, which evaluates $H_i$ at selected thresholds and bounds it from below between them by Lemma~\ref{prop:density}.

\subsubsection{Bounds at a Threshold}
At a threshold $\eta$, we run Algorithm~\ref{alg:kernel} with the excess-age cost $g_\eta\triangleq(\Delta-\eta)^+$. Substituting its bracket on $J_i(g_\eta;\bar K_{\max})$ into \eqref{eq:cvar_outer} gives $\underline H(\eta)\le H_i(\eta)\le\overline H(\eta)$, where $\underline H(\eta)\triangleq\eta+\underline J(g_\eta)/(1-\alpha)$ and $\overline H(\eta)\triangleq\eta+\overline J(g_\eta)/(1-\alpha)$. The returned budget-feasible policy $\hat\pi$ has CVaR at most $F_{\hat\pi}(\eta)=\overline H(\eta)$ by \eqref{eq:cvar_ru}. We keep the policy with the smallest such bound as $\pi_{\mathrm{save}}$ and denote this bound by $C_{\mathrm{up}}$, an upper bound on $C^\star_{\bar K_{\max},\alpha,i}$.

\subsubsection{Lower Bounds Between Evaluations}
Lemma~\ref{prop:density} bounds the density $f_\pi$ of $\mu_\pi$ by $\bar r_i$ for every budget-feasible $\pi$, and hence bounds how fast the slope $F_\pi'$ of Section~\ref{subsec:reduction} can change: $F_\pi'$ is absolutely continuous with $F_\pi''=f_\pi/(1-\alpha)\le M\triangleq\bar r_i/(1-\alpha)$ a.e. Every $F_\pi(\eta)-M\eta^2/2$ is therefore concave, and so is $H_i(\eta)-M\eta^2/2=\inf_\pi\{F_\pi(\eta)-M\eta^2/2\}$, because an infimum of concave functions is concave. Hence, on any interval $[a,b]$, $H_i$ lies above the chord through its endpoint values minus a quadratic term. Since the chord weights the endpoint values nonnegatively, we may replace them by their lower bounds and obtain, for $0\le s\le b-a$,
\begin{equation}\label{eq:chord_sag}
H_i(a+s)\ge\underline H(a)+\frac{\underline H(b)-\underline H(a)}{b-a}\,s-\frac{M}{2}\,s(b-a-s).
\end{equation}
The quadratic term, which bounds how far $H_i$ can fall below the chord, peaks at the midpoint at $M(b-a)^2/8$, so the two endpoint evaluations bound $H_i$ on the whole interval. Tightening \eqref{eq:chord_sag} with $H_i(\eta)\ge\eta$ and the range $[-\alpha/(1-\alpha),1]$ of every slope $F_\pi'$ yields a lower bound $\mathrm{LB}(a,b)$ on $\min_{\eta\in[a,b]}H_i(\eta)$, given in \eqref{eq:interval_lb_full} of the supplementary material, with $\mathrm{LB}(a,b)\ge\min\{\underline H(a),\underline H(b)\}-M(b-a)^2/8$.

\subsubsection{Interval Search and Guarantee}
We search only the thresholds between $\alpha/\bar r_i$ and $U\triangleq\operatorname{CVaR}_\alpha(\tilde\Delta_{\pi_{\mathrm{ref}}})$, where the input policy $\pi_{\mathrm{ref}}$ is budget-feasible with finite CVaR. By Lemma~\ref{prop:density}, every budget-feasible policy has VaR at least $\alpha/\bar r_i$, and its $F_\pi$ is nonincreasing below this minimizer, so $H_i$ is nonincreasing for every $\eta\le\alpha/\bar r_i$. Moreover, $U\ge\operatorname{VaR}_\alpha(\tilde\Delta_{\pi_{\mathrm{ref}}})\ge\alpha/\bar r_i$ and $C^\star_{\bar K_{\max},\alpha,i}\le U$, so $H_i(\eta)\ge\eta>U$ rules out every $\eta>U$. \textsc{Interval-Tailor} (Algorithm~\ref{alg:cvar_search}) first evaluates any $n_0\ge2$ distinct thresholds in $[\alpha/\bar r_i,U]$, including both endpoints. Adjacent evaluated thresholds form intervals, and the smallest of their lower bounds, $C_{\mathrm{lo}}$, satisfies $C_{\mathrm{lo}}\le C^\star_{\bar K_{\max},\alpha,i}\le C_{\mathrm{up}}$. Each evaluation runs Algorithm~\ref{alg:kernel} with $\delta=(1-\alpha)\gamma\epsilon$ for a fixed $\gamma\in(0,1)$, so that $\overline H(\eta)-\underline H(\eta)\le\gamma\epsilon$ by Lemma~\ref{lem:bisection_gap}. The pointwise uncertainty thus contributes at most $\gamma\epsilon$, leaving $(1-\gamma)\epsilon$ for the possible drop between evaluated thresholds: every interval of length $d$ satisfies $C_{\mathrm{up}}-\mathrm{LB}\le\gamma\epsilon+Md^2/8$. Each iteration bisects an interval attaining $C_{\mathrm{lo}}$ and evaluates its midpoint, which may lower $C_{\mathrm{up}}$, until $C_{\mathrm{up}}-C_{\mathrm{lo}}\le\epsilon$. Each half has a quarter of its parent's drop, so the bounds tighten where the optimum may lie, and an interval with $\mathrm{LB}>C_{\mathrm{up}}$ is never refined.

\begin{algorithm}[t]
\caption{\textsc{Interval-Tailor} for CVaR}
\label{alg:cvar_search}
\KwIn{Model $i$, $\alpha$, $\bar K_{\max}$, tolerance $\epsilon>0$, $\gamma\in(0,1)$, number $n_0\ge2$ of initial thresholds, and a budget-feasible policy $\pi_{\mathrm{ref}}$ with finite CVaR.}
\KwOut{A budget-feasible policy $\pi_{\mathrm{save}}$ and a bracket on $C^\star_{\bar K_{\max},\alpha,i}$ that bounds its gap.}
$U\leftarrow\operatorname{CVaR}_\alpha(\tilde\Delta_{\pi_{\mathrm{ref}}})$, $C_{\mathrm{up}}\leftarrow U$, and $\pi_{\mathrm{save}}\leftarrow\pi_{\mathrm{ref}}$\;
\textbf{Evaluate}$(\eta)$: run Algorithm~\ref{alg:kernel} on $g_\eta$ with $\delta=(1-\alpha)\gamma\epsilon$ and compute $\underline H(\eta)$ and $\overline H(\eta)$; if $\overline H(\eta)<C_{\mathrm{up}}$, set $C_{\mathrm{up}}\leftarrow\overline H(\eta)$ and $\pi_{\mathrm{save}}\leftarrow\hat\pi$\;
\textbf{Evaluate} at $n_0$ distinct thresholds of $[\alpha/\bar r_i,U]$ that include both endpoints, and compute the interval bounds and $C_{\mathrm{lo}}$\;
\While{$C_{\mathrm{up}}-C_{\mathrm{lo}}>\epsilon$}{
Bisect the interval with the smallest $\mathrm{LB}$, \textbf{Evaluate} at its midpoint, and update the bounds and $C_{\mathrm{lo}}$\;
}
\Return $\pi_{\mathrm{save}}$ and $[C_{\mathrm{lo}},C_{\mathrm{up}}]$\;
\end{algorithm}

\begin{theorem}[Guarantee of \textsc{Interval-Tailor}]
\label{prop:cvar_guarantee}
On the finite model, after initialization, the saved policy $\pi_{\mathrm{save}}$ of Algorithm~\ref{alg:cvar_search} is budget-feasible and its CVaR exceeds $C^\star_{\bar K_{\max},\alpha,i}$ by at most $C_{\mathrm{up}}-C_{\mathrm{lo}}$. The search stops after $O\big(n_0+(U-\alpha/\bar r_i)\sqrt{M/((1-\gamma)\epsilon)}\big)$ calls of Algorithm~\ref{alg:kernel}.
\end{theorem}
\begin{proof}
See Appendix~\ref{app:cvar_proof} in the supplementary material.
\end{proof}
The rate follows in three steps. First, an interval of length $d\le d_\epsilon\triangleq\sqrt{8(1-\gamma)\epsilon/M}$ falls at most $(1-\gamma)\epsilon$ below its endpoint bounds, so it cannot keep the gap above $\epsilon$. Second, while $C_{\mathrm{up}}-C_{\mathrm{lo}}>\epsilon$, the selected interval is therefore longer than $d_\epsilon$, and intervals longer than $d_\epsilon$ can be bisected only $O\big((U-\alpha/\bar r_i)/d_\epsilon\big)$ times. Third, Lemma~\ref{lem:bisection_gap} bounds each call of Algorithm~\ref{alg:kernel} by $O(\log(1/\epsilon))$ solves of \textsc{Tailor}. For fixed $n_0$ and $\gamma$, \textsc{Interval-Tailor} thus uses $O\big(\sqrt{M/\epsilon}\,\log(1/\epsilon)\big)$ solves of \textsc{Tailor} in total. The service distribution enters the call count only through $U$, as $M$ depends only on $\alpha$ and $\bar r_i$.

\section{Simulation Results}
\label{sec:simulations}
We evaluate \textsc{Tailor} under heavy-tailed delays around three questions: $i$) whether preemption merely mitigates delay variability or can exploit it; $ii$) how much AoI-dependent decisions add to the best AoI-independent preemption; and $iii$) when sampling at will pays off. We introduce the benchmarks first and answer these questions in the rest of this section.

\subsection{Benchmarks}

We compare \textsc{Tailor} with the following benchmark classes, each imposing a distinct structural restriction on the policy space. Within each class, we optimize under the same objective and budget while allowing random mixtures as in \eqref{eq:budget_mixture}, so the gap to \textsc{Tailor} measures the cost of that restriction:
\begin{itemize}
\item \emph{Zero-Wait} (GaW) samples a new update as soon as the previous one is delivered and never preempts.
\item \emph{Update-or-Wait}~\cite{sun2017update} (GaW) optimizes the waiting time before each sample but never preempts.
\item \emph{Never-Preempt} (random arrivals) discards every busy arrival, and \emph{Always-Preempt} restarts service with it.
\item \emph{Static Preemption}~\cite{arafa2019preemption,akar2026squeezingpreemption} fixes an AoI-independent service-age threshold, beyond which it preempts at the next busy arrival under random arrivals and immediately under GaW, where it also optimizes the wait.
\item \emph{Probabilistic Preemption}~\cite{moltafet2025probabilistic} preempts at each busy arrival with a fixed probability, independently of the service age and the AoI.
\end{itemize}
For the tail problems, we also optimize Static Preemption on the same grid as \textsc{Tailor} to isolate the gain from AoI-dependent decisions. The computed values match closed-form renewal values and a continuous-time Monte Carlo simulator.

\subsection{Value of Preemption}

Fig.~\ref{fig:Preemption} shows the average cost under GaW as the delay variability grows at a fixed mean. Without preemption, variability is a liability: the Zero-Wait cost grows with the second moment of the delay, so it increases exponentially in $\operatorname{Var}(\ln Y)$ under log-normal service and diverges as the Pareto~II shape approaches $a=2$; optimal waiting slows this growth but cannot stop it. With preemption, variability becomes an asset: the cost of \textsc{Tailor} decreases as the variance grows, reaching $1.45$ at $\operatorname{Var}(\ln Y)=20$, and remains finite even for $1<a\le2$. In both delay families, a larger variance at a fixed mean concentrates more mass near zero and stretches the upper tail. A non-preemptive controller must wait out every slow realization, whereas a preemptive one keeps the fast realizations and terminates the slow ones, and thus sees an effectively faster channel.

\subsection{Value of AoI-Aware Preemption}
\label{subsec:aoi_matters}

\begin{figure}[!t]
\centering
\includegraphics[width=\columnwidth]{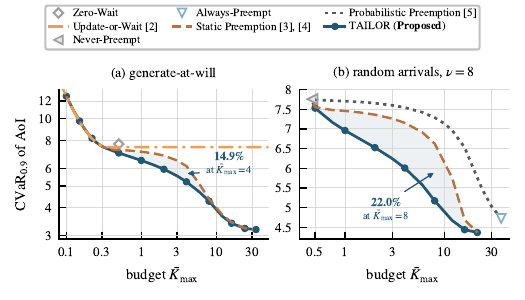}
\caption{CVaR$_{0.9}$ of the AoI versus the budget for two-scale service with $(p,r)=(0.2,20)$, $\mathbb E[Y]=2$, and $(\kappa_s,\kappa_p)=(1,5)$. Update-or-Wait, Static Preemption, and Probabilistic Preemption are optimized for the CVaR under each budget; each other baseline is a single point at its own action-cost rate.}
\label{fig:tail_budget}
\end{figure}

We consider a two-scale service time: with probability $p$, $\ln Y\sim\mathcal N(\ln m_{\mathrm f},0.5^2)$, and otherwise $\ln Y\sim\mathcal N(\ln(r m_{\mathrm f}),0.5^2)$, where $m_{\mathrm f}$ is set so that $\mathbb E[Y]=2$. Such delays arise when an update either traverses a fast path or falls back to a much slower one. Fig.~\ref{fig:tail_budget} traces the CVaR against the budget, with random arrivals at rate $\nu=8$. Update-or-Wait spends the budget only on sampling, so its CVaR stops improving once the budget exceeds the sampling rate of its unconstrained optimum, whereas \textsc{Tailor} keeps converting budget into a lower tail through preemption. The shaded gap between \textsc{Tailor} and Static Preemption is what AoI-dependent decisions add; it is small at both ends of the budget range and peaks in between, at about $15\%$ under GaW and $22\%$ under random arrivals. We explain this shape by the cost of restarts. A fresh attempt enters the fast mode with probability only $p=0.2$, so repeated restarts can incur substantial action costs before a delivery. A small budget allows few restarts in any state, and a large budget allows them in every state; in both regimes, knowing the AoI changes little, and the two frontiers nearly merge. An intermediate budget funds restarts in only some delivery cycles. \textsc{Tailor} uses these restarts in cycles that start with a large AoI and contribute most to the tail, whereas Static Preemption applies the same service-age threshold regardless of the AoI. Under random arrivals, \textsc{Tailor} matches the CVaR of Always-Preempt at about a third of its action cost. Probabilistic Preemption lies above Static Preemption at every budget: a fixed preemption probability ignores the elapsed service time, so moderate preemption probabilities even raise the CVaR above that of Never-Preempt.

\begin{figure}[!t]
\centering
\includegraphics[width=\columnwidth]{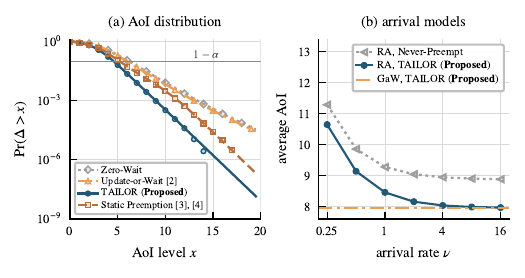}
\caption{(a) AoI distribution of the CVaR-optimal policies under GaW for two-scale service with $(p,r)=(0.2,20)$ and $\bar K_{\max}=3$. Lines show lattice values for \textsc{Tailor} and Static Preemption and exact renewal values for Zero-Wait and Update-or-Wait; markers show continuous-time Monte Carlo simulation. (b) Average AoI under random arrivals (RA) versus the arrival rate $\nu$ for bimodal service with $\kappa_s=\kappa_p=0$, together with the GaW optimum for the same service.}
\label{fig:mc_ccdf}
\end{figure}

Fig.~\ref{fig:mc_ccdf}(a) shows where the gain comes from at $(p,r)=(0.2,20)$ and $\bar K_{\max}=3$. Static Preemption selects at time zero either a short preemption age, which restarts slow attempts, or a longer one, which allows more attempts to finish. \textsc{Tailor} uses the same two regimes but selects them by the AoI: it lets the service run while the AoI is small and switches to fast restarts once the AoI is large. Its AoI distribution therefore decays faster beyond the VaR, and continuous-time Monte Carlo simulation of both policies reproduces the lattice distributions. The distributions of Zero-Wait and Update-or-Wait nearly coincide, because waiting delays the next sample but cannot shorten a slow attempt, which must run to completion and therefore governs the tail decay. Preemption changes this tail behavior: at the AoI level $15$, \textsc{Tailor} lowers the tail probability of Zero-Wait by over two orders of magnitude.

Fig.~\ref{fig:mc_ccdf}(b) compares the two arrival models under bimodal service, where $\ln Y$ follows $\mathcal N(0.75,0.75^2)$ or $\mathcal N(2,0.25^2)$ with probability $1/2$ each. A GaW controller can emulate any random-arrival policy by sampling at the epochs of an independent Poisson process, so the GaW optimum lower-bounds the random-arrival optimum at every $\nu$. With free actions ($\kappa_s=\kappa_p=0$), as $\nu$ grows, a fresh update arrives almost whenever the controller would sample or preempt, so random arrivals approach GaW and the gap closes. Under sparse arrivals, the controller must instead leave the server idle or keep an aging update in service, and the freedom to sample matters most.

\section{{Conclusion}}\label{conclusion}
We treated preemption as a control action and developed a unified framework that optimizes preemption under random arrivals, and sampling and preemption under GaW, for both the average AoI and its tail under general service times. In both models, optimal preemption reverses the role of delay variability. Without preemption, the average cost may diverge under infinite delay variance, whereas the optimal persist-or-preempt rule keeps the AoI finite whenever the mean delay is finite; under log-normal delays, its cost can even fall as the variance grows. The two models differ in what adds value: AoI-dependent preemption matters most under random arrivals with an intermediate budget, whereas the freedom to sample matters mainly when arrivals are sparse.

We establish two structural results. First, we prove that constant preemption thresholds are optimal under exponential delays, whereas general delays require state-dependent switching boundaries. Second, we show that the VaR and the CVaR require no state augmentation: a scalar threshold turns each into an average cost. In practice, cancellation hooks such as 5G NR SDU discard timers~\cite{3gpp38323} and inference preemption~\cite{vllmPreemption} can be driven by \textsc{Tailor} instead of static rules. Under heavy-tailed delays, deciding when to abandon an update in service is as fundamental to freshness as deciding when to send one.

\bibliographystyle{IEEEtran}
	\bibliography{reference}

@inproceedings{banerjee_ulukus_isit24,
  author    = {S. Banerjee and S. Ulukus},
  title     = {When to preempt in a status update system?},
  booktitle = {Proc. IEEE ISIT},
  year      = {2024},
  pages     = {1379--1384},
  doi       = {10.1109/ISIT57864.2024.10619318}
}

@ARTICLE{11393542,
  author={Li, Aimin and Wu, Shaohua and Lee, Gary C. F. and Sun, Sumei},
  journal={IEEE Trans. Inf. Theory}, 
  title={From Freshness to Effectiveness: Goal-Oriented Sampling for Remote Decision Making}, 
  year={2026},
  volume={72},
  number={4},
  pages={2237-2276}}

@article{sun2017update,
  author  = {Y. Sun and E. Uysal and R. D. Yates and C. E. Koksal and N. B. Shroff},
  title   = {Update or wait: How to keep your data fresh},
  journal = {IEEE Trans. Inf. Theory},
  volume  = {63},
  number  = {11},
  pages   = {7492--7508},
  year    = {2017}
}

@article{uysal2022semantic,
  author  = {E. Uysal and O. Kaya and A. Ephremides and J. Gross and M. Codreanu and P. Popovski and M. Assaad and G. Liva and A. Munari and B. Soret and others},
  title   = {Semantic communications in networked systems: A data significance perspective},
  journal = {IEEE Netw.},
  volume  = {36},
  number  = {4},
  pages   = {233--240},
  year    = {2022}
}

@article{li-goal,
  author  = {A. Li and S. Wu and S. Meng and R. Lu and S. Sun and Q. Zhang},
  title   = {Toward goal-oriented semantic communications: New metrics, framework, and open challenges},
  journal = {IEEE Wireless Commun.},
  volume  = {31},
  number  = {5},
  pages   = {238--245},
  year    = {2024},
  doi     = {10.1109/MWC.014.2300492}
}

@article{bensoussan1975nouvelles,
  author  = {A. Bensoussan and J.-L. Lions},
  title   = {Nouvelles m{\'e}thodes en contr{\^o}le impulsionnel},
  journal = {Appl. Math. Optim.},
  volume  = {1},
  number  = {4},
  pages   = {289--312},
  year    = {1975}
}

@book{shortle2018fundamentals,
  author    = {J. F. Shortle and J. M. Thompson and D. Gross and C. M. Harris},
  title     = {Fundamentals of queueing theory},
  publisher = {John Wiley \& Sons},
  year      = {2018}
}

@article{stettner2022approximation,
  author  = {L. Stettner},
  title   = {On an approximation of average cost per unit time impulse control of {Markov} processes},
  journal = {SIAM J. Control Optim.},
  volume  = {60},
  number  = {4},
  pages   = {2115--2131},
  year    = {2022}
}

@article{sun2019wiener,
  author  = {Y. Sun and Y. Polyanskiy and E. Uysal},
  title   = {Sampling of the {W}iener process for remote estimation over a channel with random delay},
  journal = {IEEE Trans. Inf. Theory},
  volume  = {66},
  number  = {2},
  pages   = {1118--1135},
  year    = {2020}
}

@article{TangCWYT23,
  author  = {H. Tang and Y. Chen and J. Wang and P. Yang and L. Tassiulas},
  title   = {Age optimal sampling under unknown delay statistics},
  journal = {IEEE Trans. Inf. Theory},
  volume  = {69},
  number  = {2},
  pages   = {1295--1314},
  year    = {2023}
}

@article{bedewy2021optimal,
  author  = {A. M. Bedewy and Y. Sun and S. Kompella and N. B. Shroff},
  title   = {Optimal sampling and scheduling for timely status updates in multi-source networks},
  journal = {IEEE Trans. Inf. Theory},
  volume  = {67},
  number  = {6},
  pages   = {4019--4034},
  year    = {2021}
}

@inproceedings{li2024sampling,
  author    = {A. Li and S. Wu and G. C. F. Lee and X. Chen and S. Sun},
  title     = {Sampling to achieve the goal: An age-aware remote {Markov} decision process},
  booktitle = {Proc. IEEE ITW},
  year      = {2024},
  pages     = {121--126}
}

@article{liyanaarachchi2024role,
  author  = {S. Liyanaarachchi and S. Ulukus},
  title   = {The role of early sampling in age of information minimization in the presence of {ACK} delays},
  journal = {IEEE Trans. Inf. Theory},
  volume  = {70},
  number  = {9},
  pages   = {6665--6678},
  year    = {2024}
}

@inproceedings{atasayar2025fresh,
  author    = {A. U. Atasayar and A. Li and {\c{C}}. Ar{\i} and E. Uysal},
  title     = {Fresh data delivery: Joint sampling and routing for minimizing the age of information},
  booktitle = {Proc. {ACM} MobiHoc},
  year      = {2025},
  pages     = {291--300}
}

@article{pan2023sampling,
  author    = {J. Pan and Y. Sun and N. B. Shroff},
  title     = {Sampling for remote estimation of the {W}iener process over an unreliable channel},
  journal   = {Proc. {ACM} Meas. Anal. Comput. Syst.},
  volume    = {7},
  number    = {3},
  note      = {Art. no. 60, 41 pages},
  year      = {2023}
}

@inproceedings{arafa2019preemption,
  author    = {A. Arafa and R. D. Yates and H. V. Poor},
  title     = {Timely cloud computing: Preemption and waiting},
  booktitle = {Proc. Allerton},
  year      = {2019},
  pages     = {528--535},
  doi       = {10.1109/ALLERTON.2019.8919891}
}

@article{yates2021age,
  author  = {R. D. Yates and Y. Sun and D. R. Brown and S. K. Kaul and E. Modiano and S. Ulukus},
  title   = {Age of information: An introduction and survey},
  journal = {IEEE J. Sel. Areas Commun.},
  volume  = {39},
  number  = {5},
  pages   = {1183--1210},
  year    = {2021}
}

@inproceedings{farazi2018harvesting_preemption,
  author    = {S. Farazi and A. G. Klein and D. R. Brown},
  title     = {Age of information in energy harvesting status update systems: When to preempt in service?},
  booktitle = {Proc. IEEE ISIT},
  year      = {2018},
  pages     = {2436--2440},
  doi       = {10.1109/ISIT.2018.8437904}
}

@article{wang2019preemption,
  author  = {B. Wang and S. Feng and J. Yang},
  title   = {When to preempt? {Age of Information} minimization under link capacity constraint},
  journal = {J. Commun. Netw.},
  volume  = {21},
  number  = {3},
  pages   = {220--232},
  year    = {2019},
  doi     = {10.1109/JCN.2019.000036}
}

@INPROCEEDINGS{aiminpreemption,
  author={Li, Aimin and İnce, Yiğit and Uysal, Elif},
  booktitle={Proc. IEEE ISIT}, 
  title={Taming the Heavy Tail: Age-Optimal Preemption}, 
  year={2026},
  volume={},
  number={},
  pages={1-6},
  doi={10.1109/ISIT62367.2026.11653616}}

@article{uysal_li_goal,
  author = {E. Uysal and A. Li},
  title = {Goal-oriented communications for interplanetary and non-terrestrial networks},
  journal = {IEEE Commun. Mag.},
  year = {2026},
  note = {early access}
}

@ARTICLE{10562359,
  author={Li, Aimin and Wu, Shaohua and Sun, Sumei and Cao, Jie},
  journal={IEEE Trans. Commun.}, 
  title={Goal-Oriented Tensor: Beyond Age of Information Toward Semantics-Empowered Goal-Oriented Communications}, 
  year={2024},
  volume={72},
  number={12},
  pages={7689-7704}}

@article{davis1984piecewise,
  author  = {M. H. A. Davis},
  title   = {Piecewise-deterministic {Markov} processes: A general class of non-diffusion stochastic models},
  journal = {J. R. Stat. Soc. Ser. B (Methodol.)},
  volume  = {46},
  number  = {3},
  pages   = {353--376},
  year    = {1984}
}

@inproceedings{chen2024optimal,
  author    = {X. Chen and A. Li and S. Wu},
  title     = {Optimal sampling for uncertainty-of-information minimization in a remote monitoring system},
  booktitle = {Proc. IEEE ITW},
  year      = {2024},
  pages     = {115--120}
}

@inproceedings{li2025optimal,
  author    = {A. Li and E. Uysal},
  title     = {Optimal sampling and scheduling for remote fusion estimation of correlated {W}iener processes},
  booktitle = {Proc. Annu. Allerton Conf. Commun., Control, Comput. (Allerton)},
  year      = {2025}
}

@inproceedings{Gammaawakening,
  author    = {E. Najm and R. Nasser},
  title     = {{Age of Information}: The {Gamma} awakening},
  booktitle = {Proc. IEEE ISIT},
  year      = {2016},
  pages     = {2574--2578},
  doi       = {10.1109/ISIT.2016.7541764}
}

@inproceedings{kaul2012status,
  author    = {S. K. Kaul and R. D. Yates and M. Gruteser},
  title     = {Status updates through queues},
  booktitle = {Proc. CISS},
  year      = {2012},
  pages     = {1--6}
}

@article{10559951,
  author  = {M. K. C. Shisher and Y. Sun and I.-H. Hou},
  title   = {Timely communications for remote inference},
  journal = {IEEE/{ACM} Trans. Netw.},
  volume  = {32},
  number  = {5},
  pages   = {3824--3839},
  year    = {2024},
  doi     = {10.1109/TNET.2024.3408673}
}

@inproceedings{najm2017status,
  author    = {E. Najm and R. D. Yates and E. Soljanin},
  title     = {Status updates through {M/G/1/1} queues with {HARQ}},
  booktitle = {Proc. IEEE ISIT},
  year      = {2017},
  pages     = {131--135}
}

@article{costa2016age,
  author  = {M. Costa and M. Codreanu and A. Ephremides},
  title   = {On the {Age of Information} in status update systems with packet management},
  journal = {IEEE Trans. Inf. Theory},
  volume  = {62},
  number  = {4},
  pages   = {1897--1910},
  year    = {2016}
}

@inproceedings{10619298,
  author    = {{\c{C}}. Ar{\i} and M. K. C. Shisher and E. Uysal and Y. Sun},
  title     = {Goal-oriented communications for remote inference under two-way delay with memory},
  booktitle = {Proc. IEEE ISIT},
  year      = {2024},
  pages     = {1179--1184},
  doi       = {10.1109/ISIT57864.2024.10619298}
}

@article{soysal2021age,
  author  = {A. Soysal and S. Ulukus},
  title   = {{Age of Information} in {G/G/1/1} systems: Age expressions, bounds, special cases, and optimization},
  journal = {IEEE Trans. Inf. Theory},
  volume  = {67},
  number  = {11},
  pages   = {7477--7489},
  year    = {2021}
}

@inproceedings{chen2016age,
  author    = {K. Chen and L. Huang},
  title     = {Age-of-information in the presence of error},
  booktitle = {Proc. IEEE ISIT},
  year      = {2016},
  pages     = {2579--2583}
}

@article{kosta2017age,
  author  = {A. Kosta and N. Pappas and V. Angelakis},
  title   = {{Age of Information}: A new concept, metric, and tool},
  journal = {Found. Trends Netw.},
  volume  = {12},
  number  = {3},
  pages   = {162--259},
  year    = {2017}
}

@article{asvadi2025delayed,
  author  = {S. Asvadi and F. Ashtiani},
  title   = {Delayed preemption: A new policy for balancing the {Age of Information} in prioritized networks},
  journal = {IEEE Trans. Commun.},
  year={2025},
  volume={73},
  number={12},
  pages={14639-14652}
}

@inproceedings{kaul2018age,
  author    = {S. K. Kaul and R. D. Yates},
  title     = {{Age of Information}: Updates with priority},
  booktitle = {Proc. IEEE ISIT},
  year      = {2018},
  pages     = {2644--2648}
}

@article{gunduz2023timely,
  author  = {D. G{\"u}nd{\"u}z and F. Chiariotti and K. Huang and A. E. Kal{\o}r and S. Kobus and P. Popovski},
  title   = {Timely and massive communication in {6G}: Pragmatics, learning, and inference},
  journal = {IEEE BITS Inf. Theory Mag.},
  volume  = {3},
  number  = {1},
  pages   = {27--40},
  year    = {2023}
}

@book{lawless2011statistical,
  author    = {J. F. Lawless},
  title     = {Statistical models and methods for lifetime data},
  publisher = {John Wiley \& Sons},
  year      = {2011}
}

@article{ling2022age,
  title={Age-of-information minimization in healthcare {IoT} using distributionally robust optimization},
  author={Ling, Zhuang and Hu, Fengye and Zhang, Hongliang and Han, Zhu},
  journal={IEEE Internet of Things Journal},
  volume={9},
  number={17},
  pages={16154--16167},
  year={2022},
  publisher={IEEE}
}

@article{xiao2024statistical,
  title={Statistical age of information: A risk-aware metric and its applications in status updates},
  author={Xiao, Yuquan and Du, Qinghe and Karagiannidis, George K},
  journal={IEEE Transactions on Wireless Communications},
  volume={24},
  number={3},
  pages={2325--2340},
  year={2025},
  publisher={IEEE}
}

@inproceedings{huang2026conditional,
  title={Conditional age-at-risk for task assignment across heterogeneous servers},
  author={Huang, Haoyang and Shao, Hui and Wang, Zhibo and Zhang, Meng},
  booktitle={Proc. IEEE INFOCOM},
  pages={1--10},
  year={2026},
  organization={IEEE}
}

@article{miridakis2022extreme,
  title={Extreme age of information for wireless-powered communication systems},
  author={Miridakis, Nikolaos I and Shi, Zheng and Tsiftsis, Theodoros A and Yang, Guanghua},
  journal={IEEE Wireless Communications Letters},
  volume={11},
  number={4},
  pages={826--830},
  year={2022},
  publisher={IEEE}
}

@inproceedings{zhou2020risk,
  title={Risk-aware optimization of age of information in the Internet of Things},
  author={Zhou, Bo and Saad, Walid and Bennis, Mehdi and Popovski, Petar},
  booktitle={Proc. IEEE ICC},
  pages={1--6},
  year={2020},
  organization={IEEE}
}

@inproceedings{erbayat2025corr_preempt_aoi,
  title={Age of information optimization with preemption strategies for correlated systems},
  author={Erbayat, Egemen and Maatouk, Ali and Zou, Peng and Subramaniam, Suresh},
  booktitle={Proc. IEEE ISIT},
  pages={1--6},
  year={2025},
  organization={IEEE}
}

@article{akar2026squeezingpreemption,
      title={Squeezing the Most Out of Preemption for {AoI} Minimization: Single-source Case}, 
      author={Nail Akar and Mohammad Moltafet and Sennur Ulukus and Marian Codreanu and Roy D. Yates},
      year={2026},
      journal={arXiv preprint arXiv:2607.19216}
}

@article{moltafet2026elapsed,
      title={Elapsed-Service-Time-Aware Preemption for Age of Information Minimization},
      author={Mohammad Moltafet and Nail Akar and Sennur Ulukus and Marian Codreanu and Roy D. Yates},
      year={2026},
      journal={arXiv preprint arXiv:2611.07151}
}

@article{rockafellar2000cvar,
  title={Optimization of conditional value-at-risk},
  author={Rockafellar, R Tyrrell and Uryasev, Stanislav},
  journal={Journal of risk}, number={3},
  volume={2},
  pages={21--42},
  year={2000}
}

@book{altman1999constrained,
  title={Constrained Markov decision processes},
  author={Altman, Eitan},
  year={1999},
  publisher={CRC Press}
}

@techreport{3gpp38322,
  author      = {3GPP},
  title       = {{NR; Radio Link Control (RLC) Protocol Specification}},
  institution = {{ETSI}},
  number      = {{3GPP TS 38.322, Version 19.3.0, Release 19}},
  year        = {2026},
  month       = aug
}

@techreport{3gpp38323,
  author      = {{3GPP}},
  title       = {{NR; Packet Data Convergence Protocol (PDCP) Specification}},
  institution = {{ETSI}},
  number      = {{3GPP TS 38.323, Version 19.0.0, Release 19}},
  year        = {2025}
}

@misc{grpcCancellation,
  author       = {{gRPC Authors}},
  title        = {{Cancellation}},
  howpublished = {\url{https://grpc.io/docs/guides/cancellation/}},
  year         = {2024},
  note         = {gRPC Documentation}
}

@inproceedings{vllmPreemption,
  author    = {Woosuk Kwon and Zhuohan Li and Siyuan Zhuang and Ying Sheng
               and Lianmin Zheng and Cody Hao Yu and Joseph E. Gonzalez
               and Hao Zhang and Ion Stoica},
  title     = {Efficient Memory Management for Large Language Model Serving
               with {PagedAttention}},
  booktitle = {Proceedings of the 29th Symposium on Operating Systems
               Principles (SOSP)},
  pages     = {611--626},
  year      = {2023},
  doi       = {10.1145/3600006.3613165}
}

@article{ros2Actions,
  author  = {Steven Macenski and Tully Foote and Brian Gerkey
             and Chris Lalancette and William Woodall},
  title   = {Robot Operating System 2: Design, Architecture, and Uses in the Wild},
  journal = {Science Robotics},
  volume  = {7},
  number  = {66},
  pages   = {eabm6074},
  year    = {2022},
  doi     = {10.1126/scirobotics.abm6074}
}

@book{puterman1994markov,
  author    = {M. L. Puterman},
  title     = {Markov Decision Processes: Discrete Stochastic Dynamic Programming},
  publisher = {Wiley},
  address   = {New York, NY, USA},
  year      = {1994}
}

@article{beutler1985optimal,
  author  = {F. J. Beutler and K. W. Ross},
  title   = {Optimal policies for controlled {M}arkov chains with a constraint},
  journal = {J. Math. Anal. Appl.},
  volume  = {112},
  number  = {1},
  pages   = {236--252},
  year    = {1985}
}

@article{odoni1969finding,
  author  = {A. R. Odoni},
  title   = {On finding the maximal gain for {M}arkov decision processes},
  journal = {Oper. Res.},
  volume  = {17},
  number  = {5},
  pages   = {857--860},
  year    = {1969}
}

@article{Inoue2019general,
  author  = {Yoshiaki Inoue and Hiroyuki Masuyama and Tetsuya Takine and Toshiyuki Tanaka},
  title   = {A General Formula for the Stationary Distribution of the Age of Information and Its Application to Single-Server Queues},
  journal = {IEEE Trans. Inf. Theory},
  volume  = {65},
  number  = {12},
  pages   = {8305--8324},
  month   = dec,
  year    = {2019}
}

@inproceedings{Champati2019distribution,
  author    = {Jaya Prakash Champati and Hussein Al-Zubaidy and James Gross},
  title     = {On the Distribution of {AoI} for the {GI/GI/1/1} and {GI/GI/1/2*} Systems: Exact Expressions and Bounds},
  booktitle = {Proc. IEEE INFOCOM},
  pages     = {37--45},
  month     = apr,
  year      = {2019}
}

@article{Champati2021statistical,
  author  = {Jaya Prakash Champati and Hussein Al-Zubaidy and James Gross},
  title   = {Statistical Guarantee Optimization for {AoI} in Single-Hop and Two-Hop {FCFS} Systems With Periodic Arrivals},
  journal = {IEEE Trans. Commun.},
  volume  = {69},
  number  = {1},
  pages   = {365--381},
  month   = jan,
  year    = {2021}
}

@article{Devassy2019reliable,
  author  = {Rahul Devassy and Giuseppe Durisi and Guido Carlo Ferrante and Osvaldo Simeone and Elif Uysal},
  title   = {Reliable Transmission of Short Packets Through Queues and Noisy Channels Under Latency and Peak-Age Violation Guarantees},
  journal = {IEEE J. Sel. Areas Commun.},
  volume  = {37},
  number  = {4},
  pages   = {721--734},
  month   = apr,
  year    = {2019}
}

@article{AbdelAziz2020optimized,
  author  = {Mohamed K. Abdel-Aziz and Sumudu Samarakoon and Chen-Feng Liu and Mehdi Bennis and Walid Saad},
  title   = {Optimized Age of Information Tail for Ultra-Reliable Low-Latency Communications in Vehicular Networks},
  journal = {IEEE Trans. Commun.},
  volume  = {68},
  number  = {3},
  pages   = {1911--1924},
  month   = mar,
  year    = {2020}
}

@article{Seo2019outage,
  author  = {Jun-Bae Seo and Jinho Choi},
  title   = {On the Outage Probability of Peak Age-of-Information for {D/G/1} Queuing Systems},
  journal = {IEEE Commun. Lett.},
  volume  = {23},
  number  = {6},
  pages   = {1021--1024},
  month   = jun,
  year    = {2019}
}

@article{Liu2019taming,
  author  = {Chen-Feng Liu and Mehdi Bennis},
  title   = {Taming the Tail of Maximal Information Age in Wireless Industrial Networks},
  journal = {IEEE Commun. Lett.},
  volume  = {23},
  number  = {12},
  pages   = {2442--2446},
  month   = dec,
  year    = {2019}
}

@article{sigman1990one,
  author  = {K. Sigman},
  title   = {One-dependent regenerative processes and queues in continuous time},
  journal = {Math. Oper. Res.},
  volume  = {15},
  number  = {1},
  pages   = {175--189},
  year    = {1990}
}

@inproceedings{moltafet2025probabilistic,
  author    = {M. Moltafet and H. R. Sadjadpour and Z. Rezki and M. Codreanu and R. D. Yates},
  title     = {{AoI} in {M/G/1/1} Queues with Probabilistic Preemption},
  booktitle = {Proc. IEEE Int. Symp. Inf. Theory (ISIT)},
  address   = {Ann Arbor, MI, USA},
  year      = {2025}
}

@article{Champati2021minimum,
author = {Jaya Prakash Champati and Ramana R. Avula and Tobias J. Oechtering and James Gross},
title = {Minimum Achievable Peak Age of Information Under Service Preemptions and Request Delay},
journal = {IEEE J. Sel. Areas Commun.},
volume = {39},
number = {5},
pages = {1365--1379},
month = may,
year = {2021}
}

@article{chen2025rsmaxurllc,
  author = {Y. Chen and H. Lu and C. Wu and L. Qin and X. Guo},
  title = {Performance optimization in {RSMA}-assisted uplink {xURLLC} {IIoT} networks with statistical {QoS} provisioning},
  journal = {IEEE Trans. Wireless Commun.},
  volume = {24},
  number = {12},
  pages = {10100--10115},
  year = {2025}
}

@misc{chen2025aoioffloading,
  author = {Y. Chen and F. Guo and C. Wu and S. Liu and H. Lu and C. W. Chen},
  title = {{AoI}-aware task offloading and transmission optimization for industrial {IoT} networks: A branching deep reinforcement learning approach},
  howpublished = {arXiv:2510.16414},
  year = {2025}
}

@article{pan2026tailrisk,
  author = {H. Pan and S. Chen and Z. Wang and T.-T. Chan},
  title = {Tail-risk-sensitive scheduling of version age of information via distributional reinforcement learning},
  journal = {IEEE Trans. Mobile Comput.},
  year = {2026}
}

@misc{pan2026persistence,
  author = {H. Pan and C. Chen and S. Zhou and K. Chen and T.-T. Chan},
  title = {A persistence-aware framework for age violation control in wireless status update systems},
  howpublished = {arXiv:2605.13002},
  year = {2026}
}

@article{liu2023aos,
  author = {W. Liu and D. Li and T. Liang and T. Zhang and Z. Lin and N. Al-Dhahir},
  title = {Joint trajectory and scheduling optimization for age of synchronization minimization in {UAV}-assisted networks with random updates},
  journal = {IEEE Trans. Commun.},
  volume = {71},
  number = {11},
  pages = {6633--6646},
  year = {2023}
}

@article{zhang2025fbl,
  author = {D. Zhang and M. Sun and L. Song and S. Jia and A. Li and T. Zhang and A. Al-Dulaimi and S. Mumtaz},
  title = {Information freshness and timeliness analysis in the finite blocklength regime for mission-critical applications},
  journal = {IEEE Trans. Commun.},
  volume = {73},
  number = {12},
  pages = {14458--14468},
  year = {2025}
}

@article{xie2023importance,
  author = {X. Xie and H. Wang},
  title = {Importance-aware optimization for age of information in multi-cluster {IoT} systems with random delays},
  journal = {IEEE Commun. Lett.},
  volume = {27},
  number = {2},
  pages = {746--750},
  year = {2023}
}

@article{xie2024multisensor,
  author = {X. Xie and H. Wang and X. Liu},
  title = {Scheduling for minimizing the age of information in multisensor multiserver industrial {Internet of Things} systems},
  journal = {IEEE Trans. Ind. Informat.},
  volume = {20},
  number = {1},
  pages = {573--582},
  year = {2024}
}
    				\begin{IEEEbiographynophoto}{Aimin Li}
(Member, IEEE) received the B.S. degree (Best Thesis Award) and the Ph.D. degree (Best Dissertation Award nomination) from the Harbin Institute of Technology, Shenzhen, China, in 2020 and 2025, respectively. He was a Visiting Researcher with A*STAR I$^2$R, Singapore, from 2023 to 2024, and a Postdoctoral Researcher with Middle East Technical University, Ankara, T\"urkiye, from 2025 to 2026. He is currently a Postdoctoral Researcher with KTH Royal Institute of Technology, Stockholm, Sweden. His research interests include age of information, goal-oriented communications, and non-terrestrial networks. He is a Workshop Co-Chair of IEEE ICC 2027, a Guest Editor for \textit{Entropy} and \textit{Electronics}, and a reviewer for \textit{IEEE Trans. Netw.}, \textit{IEEE Trans. Inf. Theory}, \textit{IEEE J. Sel. Areas Commun.}, etc.
\end{IEEEbiographynophoto}

 	\begin{IEEEbiographynophoto}{Yiğit İnce}
        (Student Member, IEEE) received the B.S. degree in Electrical and Electronics Engineering from Middle East Technical University (METU) in 2025, graduating as the valedictorian. He is currently pursuing the M.Sc. degree in the same department. His research interests include random access protocols, wireless communications, non-terrestrial networks, and decision-making theory. He has served as a reviewer for \textit{IEEE Transactions on Communications}.
        \end{IEEEbiographynophoto}

		\begin{IEEEbiographynophoto}{Elif Uysal} (Fellow, IEEE) is a Professor of Electrical and Electronics Engineering at the Middle East Technical University (METU), in Ankara, Turkey. She received the Ph.D. degree from Stanford University in 2003, the S.M. degree from the Massachusetts Institute of Technology (MIT) in 1999 and the B.S. in 1997 from METU. She was elected IEEE Fellow for “pioneering contributions to energy-efficient and low-latency communications”, and Fellow of the Artificial Intelligence Industry Alliance in 2022. She was a recipient of the IEEE INFOCOM2026 Test of Time Paper Award, and the 2025 Selcuk Yasar Award. Dr. Uysal received an ERC Advanced Grant (GO SPACE) in 2024, a TUBITAK BIDEB National Pioneer Researcher Grant (2020), and the Science Academy of Turkey Young Scientist Award (2014). She has chaired the METU Parlar Foundation for Education and Research since 2022. She is the founder of FRESHDATA Technologies (2022). She has served on the editorial boards of the \textit{IEEE Transactions on Information Theory}, the \textit{IEEE/ACM Transactions on Networking}, and the \textit{IEEE Transactions on Wireless Communications}, and as Lead Guest Editor for the \textit{IEEE Journal on Selected Areas in Information Theory}.
\end{IEEEbiographynophoto}

\clearpage
\setcounter{page}{1}
\renewcommand{\thepage}{S\arabic{page}}
\setcounter{equation}{0}
\renewcommand{\theequation}{S\arabic{equation}}
\renewcommand{\theHequation}{S\arabic{equation}}
\setcounter{lemma}{0}
\renewcommand{\thelemma}{S\arabic{lemma}}
\renewcommand{\theHlemma}{S\arabic{lemma}}
\setcounter{proposition}{0}
\renewcommand{\theproposition}{S\arabic{proposition}}
\renewcommand{\theHproposition}{S\arabic{proposition}}
\raggedbottom
\markboth{IEEE Transactions on Networking: Supplementary Material}%
{Li \MakeLowercase{\textit{et al.}}: Supplementary Material}
\twocolumn[{%
\begin{center}
{\LARGE Supplementary Material for
\textquotedblleft Optimal Preemption:\\ Guaranteeing Information Freshness in Mean Age and Beyond\textquotedblright\par}
\vspace{1.5ex}
{\normalsize Aimin Li, Yi\u{g}it \.{I}nce,  and Elif Uysal\par}
\vspace{3ex}
\end{center}}]
\appendices
\renewcommand{\theHsection}{supp.\Alph{section}}

\section{{Verification of Theorems~\ref{thm:random_integral_bellman} and~\ref{thm:integral_bellman}}}
\label{app:random_integral_bellman_proof}
In Appendices~\ref{app:random_integral_bellman_proof} and~\ref{app:exp_proof}, $\kappa_s$ and $\kappa_p$ stand for $\lambda\kappa_s$ and $\lambda\kappa_p$ in Problem~\ref{p1}. By the integral dynamic programming principle, we condition on the first event after the current epoch. Let $A\sim\mathrm{Exp}(\nu)$ denote the time until the next arrival, independent of the residual service time $R_b$.

\medskip
\noindent\textit{I. Idle Phase}

An idle server admits the next arrival at cost $\kappa_s$. Since $\Pr(A>t)=e^{-\nu t}$, conditioning on $A$ gives
\begin{equation}
\begin{split}
h_I(\Delta)&=\int_0^\infty e^{-\nu t}\big(g(\Delta+t)-\rho\big)\,dt+\kappa_s\\
&\quad+\int_0^\infty\nu e^{-\nu t}h_B(\Delta+t,0)\,dt,
\end{split}
\end{equation}
which is \eqref{eq:random_h_idle}.

\medskip
\noindent\textit{II. Busy Phase}

For $0\le b\le\Delta$ with $\bar F_Y(b)>0$, let $U=\min\{A,R_b\}$. By independence, $\Pr(U>t)=e^{-\nu t}S_b(t)$, and the arrival-first and completion-first events have measures $\nu e^{-\nu t}S_b(t)\,dt$ and $e^{-\nu t}\,dF_{R_b}(t)$, respectively. A completion at time $t$ leaves the server idle with AoI $b+t$, whereas an arrival at time $t$ triggers the choice between persisting and preempting. Conditioning on $U$ gives
\begin{equation}
\begin{split}
h_B(\Delta,b)&=\int_0^\infty e^{-\nu t}S_b(t)\big(g(\Delta+t)-\rho\big)\,dt\\
&\quad+\int_{[0,\infty)}e^{-\nu t}h_I(b+t)\,dF_{R_b}(t)\\
&\quad+\int_0^\infty\nu e^{-\nu t}S_b(t)\min\big\{h_B(\Delta+t,b+t),\\
&\qquad\qquad\kappa_p+h_B(\Delta+t,0)\big\}\,dt,
\end{split}
\end{equation}
which is \eqref{eq:random_h_busy}.

\medskip
\noindent\textit{III. Preemption Rule}

At a busy arrival, persisting has value $h_B(\Delta,b)$ and preempting has value $\kappa_p+h_B(\Delta,0)$. Hence \eqref{eq:random_preempt_rule} defines a measurable deterministic minimizer $\pi^\star$ of the busy-phase ACOE, with ties resolved by preemption.

\medskip
\noindent\textit{IV. Verification}

Suppose that $\max\{|h_I(\Delta)|,|h_B(\Delta,b)|\}\le K(1+\Delta^m)$ on the reachable states for some $K<\infty$ and $m\ge0$, which, together with the growth of $g$, ensures finite-horizon integrability. Iterating the ACOEs along the arrival and completion epochs, as in \cite[App.~B]{aiminpreemption}, and truncating at a deterministic time $T$ gives, for every admissible $\pi$ and initial state $x=X(0)$,
\begin{equation}\label{eq:random_verification_bound}
h(x)\le\mathbb E_x^\pi\big[C_T^\pi-\rho T+h(X(T))\big],
\end{equation}
where $C_T^\pi\triangleq\int_0^T g(\Delta(t))\,dt+\kappa_sN_s^\pi(T)+\kappa_pN_p^\pi(T)$, with equality for $\pi^\star$. Arrivals are independent of service, and the minimizer in \eqref{eq:random_preempt_rule} is Borel and acts only at arrival epochs, so $\pi^\star$ is nonexplosive. We assume the transversality condition
\begin{equation}\label{eq:transversality}
\lim_{T\to\infty}\mathbb E_x^\pi|h(X(T))|/T=0
\end{equation}
for every admissible $\pi$ with finite average cost and for $\pi^\star$; dividing \eqref{eq:random_verification_bound} by $T$ then yields $\rho^\pi\ge\rho=\rho^{\pi^\star}$. For $g(\Delta)=\Delta$ and $m=1$, the AoI grows at unit rate and jumps only downward, so $\Delta(T)^2\le\Delta(0)^2+2\int_0^T\Delta(t)\,dt$ pathwise; a finite average cost gives $\mathbb E_x^\pi[\Delta(T)^2]=O(T)$, hence $\mathbb E_x^\pi[\Delta(T)]=O(\sqrt T)$ gives \eqref{eq:transversality}.

\medskip
\noindent\textit{V. Verification for Theorem~\ref{thm:integral_bellman}}
\label{app:gaw_verification}

Under GaW, we extend $v$ to every busy state $(\Delta,b)$ by
\begin{equation}\label{eq:gaw_HB}
\begin{split}
H_B(\Delta,b)&\triangleq\inf_{u\in[b,\infty]}\mathbb E\Big[\int_0^{\min\{R_b,u-b\}}\big(g(\Delta+t)-\rho\big)\,dt\\
&\quad+\mathbf 1_{\{R_b\le u-b\}}h_I(b+R_b)\\
&\quad+\mathbf 1_{\{R_b>u-b\}}\big(\kappa_p+v(\Delta+u-b)\big)\Big],
\end{split}
\end{equation}
where $u=\infty$ means no preemption, so $H_B(y,0)=v(y)$. Let $H_B$ be finite, Borel, and of polynomial growth, and let \eqref{eq:transversality} hold with $h=(h_I,H_B)$. The trajectory is deterministic between impulse and completion epochs, so \eqref{eq:idle_integral_bellman} and \eqref{eq:gaw_HB} give \eqref{eq:random_verification_bound} for every admissible policy, with equality for the minimizers of Theorem~\ref{thm:integral_bellman}; dividing by $T$ proves their optimality.

\section{{Proof of Theorem~\ref{thm:exp}}}\label{app:exp_proof}
As in Appendix~\ref{app:random_integral_bellman_proof}, $\kappa_s$ and $\kappa_p$ include the factor $\lambda$. For each model, we consider a finite Borel-measurable ACOE solution with $|h_I(x)|\le K(1+x^m)$ and, respectively, $|h_B(\Delta,b)|\le K(1+\Delta^m)$ on $0\le b\le\Delta$ or $|v(y)|\le K(1+y^m)$, where $m\ge\ell$. The mechanism behind both parts is the memoryless completion clock: restarting service changes the age of the packet that will be delivered, but not the law of the time until delivery, so the AoI enters the busy-phase value only through an additive term.

\medskip
\noindent\textit{I. Common Reduction}

Let $G$ be defined as in Theorem~\ref{thm:exp}, with the average cost $\rho$ of the model at hand. Splitting its defining integral at $\theta$ gives
\begin{equation}\label{eq:G_semigroup_supp}
G(x)=\int_0^\theta e^{-\beta t}\big(g(x+t)-\rho\big)\,dt+e^{-\beta\theta}G(x+\theta),
\end{equation}
which holds for finite $\theta\ge0$ and extends to $\theta=\infty$ by polynomial growth. Multiplying \eqref{eq:G_semigroup_supp} by $\nu e^{-\nu\theta}$ and integrating over $\theta$ yields, by Fubini's theorem,
\begin{equation}\label{eq:G_cancellation_random}
\begin{split}
&\int_0^\infty\nu e^{-(\nu+\beta)t}G(x+t)\,dt\\
&\quad=G(x)-\int_0^\infty e^{-(\nu+\beta)t}\big(g(x+t)-\rho\big)\,dt.
\end{split}
\end{equation}
In part (a), $R_b\sim\mathrm{Exp}(\beta)$ for every $b$, so $S_b(t)=e^{-\beta t}$ and $dF_{R_b}(t)=\beta e^{-\beta t}\,dt$ in \eqref{eq:random_h_busy}. Set $u_R(\Delta,b)\triangleq h_B(\Delta,b)-G(\Delta)$. Substituting into \eqref{eq:random_h_busy} and applying \eqref{eq:G_cancellation_random} cancels the running cost:
\begin{equation}\label{eq:u_random_fixed_point}
\begin{split}
u_R(\Delta,b)&=\int_0^\infty\beta e^{-(\nu+\beta)t}h_I(b+t)\,dt\\
&\quad+\int_0^\infty\nu e^{-(\nu+\beta)t}\min\big\{u_R(\Delta+t,b+t),\\
&\qquad\qquad\kappa_p+u_R(\Delta+t,0)\big\}\,dt.
\end{split}
\end{equation}
In part (b), set $u_G(y)\triangleq v(y)-G(y)$ and $H(\theta)\triangleq\int_0^\theta\beta e^{-\beta t}h_I(t)\,dt+e^{-\beta\theta}\kappa_p$. By \eqref{eq:busy_acoe_Q} and \eqref{eq:G_semigroup_supp}, $Q(y,\theta;\rho)=G(y)+H(\theta)+e^{-\beta\theta}u_G(y+\theta)$, so
\begin{equation}\label{eq:translation_invariant}
u_G(y)=\inf_{\theta\in[0,\infty]}\big\{H(\theta)+e^{-\beta\theta}u_G(y+\theta)\big\},
\end{equation}
where the last term is read as zero at $\theta=\infty$. In both models, the AoI now enters only through the arguments of the transformed function.

\medskip
\noindent\textit{II. Shift Invariance}

For part (a), let $\mathcal T_a$ denote the operator on the right side of \eqref{eq:u_random_fixed_point} and define $\|\varphi\|_\omega\triangleq\sup_{0\le b\le\Delta}e^{-\omega\Delta}|\varphi(\Delta,b)|$ with $\omega=\beta/2$. Since $|\min\{a_1,c_1\}-\min\{a_2,c_2\}|\le\max\{|a_1-a_2|,|c_1-c_2|\}$ and every transition stays in $0\le b\le\Delta$,
\begin{equation}\label{eq:random_yb_contraction_local}
\|\mathcal T_a\varphi-\mathcal T_a\psi\|_\omega\le q_\omega\|\varphi-\psi\|_\omega,
\quad q_\omega\triangleq\frac{\nu}{\nu+\beta-\omega}<1.
\end{equation}
Polynomial growth gives $\|u_R\|_\omega<\infty$. For $c\ge0$, the shift $u_{R,c}(\Delta,b)\triangleq u_R(\Delta+c,b)$ is again a fixed point of $\mathcal T_a$, and $\|u_{R,c}\|_\omega\le e^{\omega c}\|u_R\|_\omega$. Hence $\|u_{R,c}-u_R\|_\omega\le q_\omega\|u_{R,c}-u_R\|_\omega$, which forces $u_{R,c}=u_R$.

For part (b), assume as in Theorem~\ref{thm:integral_bellman} that the infimum in \eqref{eq:translation_invariant} is attained by a measurable selection $\theta^\star(y)\in[\epsilon,\infty]$ for every $y\ge0$ and some $\epsilon>0$. Set $u_{G,c}(y)\triangleq u_G(y+c)$ and $d_c\triangleq u_{G,c}-u_G$; both $u_G$ and $u_{G,c}$ satisfy \eqref{eq:translation_invariant}. Starting from $y_0=y$, use the minimizer $\theta_k=\theta^\star(y_k)$ for $u_G(y_k)$ and the same $\theta_k$ as a competitor for $u_{G,c}(y_k)$. Subtracting gives
\begin{equation}\label{eq:shift_sandwich_supp}
d_c(y_k)\le e^{-\beta\theta_k}d_c(y_{k+1}),\qquad y_{k+1}\triangleq y_k+\theta_k.
\end{equation}
If some $\theta_k=\infty$, then $d_c(y)\le0$ at once. Otherwise, iterating \eqref{eq:shift_sandwich_supp} gives $d_c(y)\le e^{-\beta S_n}d_c(y+S_n)$ with $S_n=\sum_{k<n}\theta_k\ge n\epsilon$, and the right side vanishes as $n\to\infty$ by polynomial growth. Hence $d_c(y)\le0$. Exchanging the roles of $u_G$ and $u_{G,c}$, with the minimizer $\theta^\star(y_k+c)$ of $u_{G,c}(y_k)$, gives $d_c(y)\ge0$. Thus $u_{G,c}=u_G$ for every $c\ge0$.

\medskip
\noindent\textit{III. Consequences}

In part (a), $u_R$ does not depend on $\Delta$, so $u_R(\Delta,b)=f(b)$ and
\begin{equation}\label{eq:random_yb_separable_supp}
h_B(\Delta,b)=G(\Delta)+f(b).
\end{equation}
The switching function in \eqref{eq:random_preempt_rule} becomes
\begin{equation}\label{eq:random_yb_switching}
\Phi_1(\Delta,b)=f(b)-f(0)-\kappa_p,
\end{equation}
which is independent of $\Delta$. In part (b), $u_G$ is constant, and $v(0)=0$ gives $u_G\equiv-G(0)$. Therefore
\begin{equation}\label{eq:gaw_exp_closed_supp}
\begin{split}
v(y)&=G(y)-G(0),\\
Q(y,\theta;\rho)&=G(y)+H(\theta)-e^{-\beta\theta}G(0),
\end{split}
\end{equation}
so the set of minimizing stopping ages, which may contain $\theta=\infty$, is the same for every $y$, and $\theta^\star$ can be chosen constant.

\medskip
\noindent\textit{IV. Constant Threshold in Part (a)}

Substituting \eqref{eq:random_yb_separable_supp} into \eqref{eq:random_h_idle} gives
\begin{equation}\label{eq:idle_value_random_decomp}
\begin{split}
h_I(\Delta)&=\kappa_s+f(0)+\int_0^\infty e^{-\nu t}\big(g(\Delta+t)-\rho\big)\,dt\\
&\quad+\int_0^\infty\nu e^{-\nu t}G(\Delta+t)\,dt.
\end{split}
\end{equation}
Since $g$ and $G$ are nondecreasing, so is $h_I$. By \eqref{eq:u_random_fixed_point}, $f$ satisfies
\begin{equation}\label{eq:f_random_fixed_point}
\begin{split}
f(b)&=\int_0^\infty e^{-(\nu+\beta)t}\big[\beta h_I(b+t)\\
&\qquad+\nu\min\{f(b+t),\kappa_p+f(0)\}\big]\,dt.
\end{split}
\end{equation}
Let $\mathcal T_1$ denote the right side of \eqref{eq:f_random_fixed_point} with $f$ replaced by $\varphi$. Under $\|\varphi\|_{\omega,1}\triangleq\sup_{b\ge0}e^{-\omega b}|\varphi(b)|$, the same argument as in Part II shows that $\mathcal T_1$ is a contraction with modulus $q_\omega$, so $f$ is its unique fixed point. Because $h_I$ is nondecreasing and $b\mapsto\min\{\varphi(b+t),\kappa_p+\varphi(0)\}$ is nondecreasing whenever $\varphi$ is, $\mathcal T_1$ preserves nondecreasing functions. Iterating from $\varphi_0\equiv0$ therefore shows that $f$ is nondecreasing. Moreover, the substitution $s=b+t$ writes $(\mathcal T_1\varphi)(b)$ as $e^{(\nu+\beta)b}$ times a tail integral over $[b,\infty)$ of a function that is integrable on $[0,\infty)$ by the weighted-norm bound, which is continuous in $b$; hence $f$ is continuous. Thus $\{b\ge0:f(b)\ge f(0)+\kappa_p\}$ is a closed upper set, and with $\inf\varnothing=\infty$,
\begin{equation}\label{eq:random_threshold_supp}
b^\star\triangleq\inf\{b\ge0:f(b)\ge f(0)+\kappa_p\}.
\end{equation}
By \eqref{eq:random_yb_switching}, preempting if and only if $b\ge b^\star$ is a Bellman-minimizing rule, which proves part (a).

\medskip
\noindent\textit{V. Sampling Threshold in Part (b)}

For $g(x)=x$, \eqref{eq:gaw_exp_closed_supp} gives $v(y)=y/\beta=y\,\mathbb E[Y]$. Substituting into \eqref{eq:idle_integral_bellman} yields the same quadratic minimization as in Corollary~\ref{cor:no_preemption},
\begin{equation}\label{eq:exp_sampling_supp}
\begin{split}
z^\star(\Delta)&=\arg\min_{z\ge\Delta}\Big\{\frac{z^2}{2}+\Big(\frac1\beta-\rho\Big)z\Big\}\\
&=\max\Big\{\Delta,\rho-\frac1\beta\Big\},
\end{split}
\end{equation}
which completes the proof.

\section{Far-Field Closure Under Random Arrivals}
\label{app:far_field_random}
Here $\kappa_s$ and $\kappa_p$ are the unweighted action costs, and we write $\lambda$ explicitly. Let $A\sim\mathrm{Exp}(\nu)$ be independent of $Y$. Under the closure policy, each attempt lasts $T\triangleq\min\{Y,A\}$ and delivers its update if $Y<A$, which occurs with probability $p_\nu=\mathbb{E}[e^{-\nu Y}]$. The time to delivery therefore satisfies $D\overset{d}{=}T+\mathbf{1}_{\{A<Y\}}D'$, where $D'$ is an independent copy of $D$. With $L_\nu\triangleq\int_0^\infty t\,e^{-\nu t}\bar F_Y(t)\,dt$, we have $\mathbb{E}[T]=(1-p_\nu)/\nu$, $\mathbb{E}[T^2]=2L_\nu$, and $\mathbb{E}[T\mathbf{1}_{\{A<Y\}}]=\nu L_\nu$, which yield
\begin{equation}\label{eq:far_moments}
\mathbb{E}[D]=\frac{1-p_\nu}{\nu p_\nu},\qquad \mathbb{E}[D^2]=\frac{2L_\nu}{p_\nu^2}.
\end{equation}
Because $T\le A$ and each attempt delivers with probability $p_\nu>0$, the geometric sum gives $\mathbb E[e^{sD}]=\mathbb E[e^{sT}\mathbf 1_{\{Y<A\}}]/(1-\mathbb E[e^{sT}\mathbf 1_{\{A<Y\}}])<\infty$ for small $s>0$, as the denominator is $p_\nu$ at $s=0$, so every moment of $D$ is finite. The number of preemptions before delivery is geometric with mean $(1-p_\nu)/p_\nu$, and the service time $W$ of the delivered update has the exponentially tilted law $\Pr\{W\in dw\}=e^{-\nu w}\,dF_Y(w)/p_\nu$.

For a far-field rule that preempts only at service ages of at least $k\in[0,\infty]$, the clock $C_k\triangleq k+A$ ($C_\infty\triangleq\infty$) replaces $A$: an attempt lasts $T_k\triangleq\min\{Y,C_k\}$ and succeeds with probability $p_k\triangleq F_Y(k)+\int_{(k,\infty)}e^{-\nu(y-k)}\,dF_Y(y)$, $D_k\overset{d}{=}T_k+\mathbf 1_{\{C_k<Y\}}D_k'$, $\Pr\{W_k\in dw\}=\Pr\{C_k>w\}\,dF_Y(w)/p_k$, and the expressions below hold with $(p_k,D_k,W_k)$ in place of $(p_\nu,D,W)$. For $k<\infty$, $D_k$ has all moments; for $k=\infty$, $D_\infty=Y$, and linear and hinge costs require $\mathbb E[Y^2]<\infty$.

With $G(x)\triangleq\mathbb{E}[\int_0^D g(x+t)\,dt]$, \eqref{eq:tail_closure_random} reads $h_{\mathrm{tail}}(x)=G(x)-\rho\,\mathbb{E}[D]+\lambda\kappa_p(1-p_\nu)/p_\nu+c$. For $g(\Delta)=\Delta$, $G(x)=x\,\mathbb{E}[D]+\mathbb{E}[D^2]/2$. For the VaR running cost, $G(x)=\mathbb{E}[D]$ for $x\ge\Delta_{\mathrm{th}}$. For the CVaR running cost, $G(x)=(x-\eta)\mathbb{E}[D]+\mathbb{E}[D^2]/2$ for $x\ge\eta$, the factor $1/(1-\alpha)$ entering only the outer objective.

At the delivery instant, the system is idle with AoI $W$, so $c=\mathbb{E}[h_I(W)]$. By \eqref{eq:random_h_idle}, $h_I(w)=J_\nu(w)-\rho/\nu+\lambda\kappa_s+\mathbb{E}[\tilde h(w+L')]$, where $J_\nu(w)\triangleq\int_0^\infty e^{-\nu t}g(w+t)\,dt$, $L'\sim\mathrm{Exp}(\nu)$ is the idle period, and $\tilde h(z)\triangleq h_B(z,0)$ for $z\le\Delta_{\mathrm{cut}}$ and $\tilde h(z)\triangleq h_{\mathrm{tail}}(z)$ otherwise. With $Z\triangleq W+L'$, this gives
\begin{equation}\label{eq:c_closure}
\begin{split}
\Pr\{Z\le\Delta_{\mathrm{cut}}\}\,c&=\mathbb{E}[J_\nu(W)]-\frac{\rho}{\nu}+\lambda\kappa_s+\mathbb{E}\big[h_B(Z,0)\mathbf{1}_{\{Z\le\Delta_{\mathrm{cut}}\}}\big]\\
&\quad+\mathbb{E}\big[\big(h_{\mathrm{tail}}(Z)-c\big)\mathbf{1}_{\{Z>\Delta_{\mathrm{cut}}\}}\big],
\end{split}
\end{equation}
Here $h_{\mathrm{tail}}(Z)-c=G(Z)-\rho\,\mathbb{E}[D]+\lambda\kappa_p(1-p_\nu)/p_\nu$ does not involve $c$. Choosing $\Delta_{\mathrm{cut}}$ with $\Pr\{Y<\Delta_{\mathrm{cut}}\}>0$ ensures $\Pr\{Z\le\Delta_{\mathrm{cut}}\}>0$, since $W<\Delta_{\mathrm{cut}}$ with positive probability and $L'$ is exponential, so \eqref{eq:c_closure} determines $c$ for given $(\rho,h_B)$. It is affine in $(\rho,h_B,c)$, so we append it to the linear evaluation system as one equation in the additional unknown $c$, rather than eliminating $c$ explicitly.

On the grid, \eqref{eq:r_pi} and \eqref{eq:P_pi} are modified only where a transition leaves the triangle. A busy arrival at $\Delta+t>\Delta_{\mathrm{cut}}$ is preempted and contributes $\lambda\kappa_p+h_{\mathrm{tail}}(\Delta+t)$, and an admission at $b+t+s>\Delta_{\mathrm{cut}}$ contributes $h_{\mathrm{tail}}(b+t+s)$, with $\lambda\kappa_s$ already counted in \eqref{eq:r_pi}. On the finite model, an arrival occurs at the end of a step with probability $1-e^{-\nu\,dt}$, and a completion in the same step is processed before the arrival; the moments in \eqref{eq:far_moments}, the delivered-age law, and $c$ are therefore evaluated by the corresponding slotted recursion, which recovers the continuous-time expressions as $dt\to0$. For $k>0$, a transition leaving the grid at service age $b>0$ conditions on $R_b$ through the completion probabilities $1-\bar F_Y((j+1)dt)/\bar F_Y(j\,dt)$. For a fixed discretization and far-field rule, enlarging the cutoff nests the restricted policy classes when the previous policy can be reproduced on the extended grid. Grid refinement and Monte Carlo simulation assess the finite model empirically.

\section{Supporting Results for Tail-Risk Optimization}
\label{app:tail_guarantees}
In this appendix, we write $\lambda$ explicitly, so $\kappa_s$ and $\kappa_p$ denote the unweighted action costs of Section~\ref{sec:model}. Appendix~\ref{app:duality_proof} and the threshold reductions work on the continuous model, whereas Lemmas~\ref{prop:kernel} and~\ref{lem:bisection_gap}, the bounds of Algorithm~\ref{alg:kernel}, and the guarantees of Algorithm~\ref{alg:var_search} and Theorem~\ref{prop:cvar_guarantee} refer to the finite model that \textsc{Tailor} solves. We consider policies for which the limit defining $\mu_\pi$ in Section~\ref{subsec:risk_formulations} is a probability measure, with finite mean only for the CVaR, and we require the average costs used in Section~\ref{sec:tail} to agree with their occupation-measure integrals, which we call cost consistency. Since an AoI path spends zero time at any fixed age, $\mu_\pi$ is atomless.

\subsection{Verification of \eqref{eq:var_outer} and \eqref{eq:cvar_outer}}
\label{app:verification}
By the definition of $\mu_\pi$, $\mathbb P(\tilde\Delta_\pi>\Delta_{\mathrm{th}})$ equals $\bar g_\pi$ under the running cost $\mathbf 1_{\{\Delta>\Delta_{\mathrm{th}}\}}$, so $J_i(\mathbf 1_{\{\Delta>\Delta_{\mathrm{th}}\}};\bar K_{\max})$ is the infimum of this probability over budget-feasible policies. Each such probability is nonincreasing in $\Delta_{\mathrm{th}}$, and hence so is $J_i(\mathbf 1_{\{\Delta>\Delta_{\mathrm{th}}\}};\bar K_{\max})$. For every feasible $\pi$, right-continuity gives $J_i(\mathbf 1_{\{\Delta>\operatorname{VaR}_\alpha(\tilde\Delta_\pi)\}};\bar K_{\max})\le\mathbb P(\tilde\Delta_\pi>\operatorname{VaR}_\alpha(\tilde\Delta_\pi))\le1-\alpha$, so $\operatorname{VaR}_\alpha(\tilde\Delta_\pi)$ belongs to the set in \eqref{eq:var_outer}; taking the infimum over $\pi$ shows that the right-hand side of \eqref{eq:var_outer} never exceeds $V^\star_{\bar K_{\max},\alpha,i}$. Conversely, suppose that $J_i(\mathbf 1_{\{\Delta>\Delta_{\mathrm{th}}\}};\bar K_{\max})\le1-\alpha$ and that a feasible $\hat\pi$ attains this infimum. Such a $\hat\pi$ exists on the finite model of Section~\ref{subsec:tailor}, where a random mixture of two deterministic stationary policies attains the optimal value of Problem~\ref{p4} (Lemma~\ref{prop:kernel}). Then $\mathbb P(\tilde\Delta_{\hat\pi}>\Delta_{\mathrm{th}})\le1-\alpha$, so $V^\star_{\bar K_{\max},\alpha,i}\le\operatorname{VaR}_\alpha(\tilde\Delta_{\hat\pi})\le\Delta_{\mathrm{th}}$; taking the infimum over such $\Delta_{\mathrm{th}}$ gives equality in \eqref{eq:var_outer} on the finite model, and on the continuous model whenever these optimizers exist. For \eqref{eq:cvar_outer}, $\mu_\pi$ is atomless, so \eqref{eq:cvar_ru} over $\eta\in\mathbb R$ and its minimizer $\operatorname{VaR}_\alpha(\tilde\Delta_\pi)\ge0$ follow from \cite[Thm.~1]{rockafellar2000cvar}, and restricting $\eta$ to $\mathbb R_+$ is lossless. By cost consistency, $\mathbb E[(\tilde\Delta_\pi-\eta)^+]$ equals $\bar g_\pi$ under the running cost $(\Delta-\eta)^+$. Exchanging the infima over $\pi$ and $\eta$ gives \eqref{eq:cvar_outer}.

\subsection{Proof of Lemma~\ref{prop:duality}}
\label{app:duality_proof}
We first construct strictly feasible policies of finite cost.
\begin{lemma}[Strictly Feasible Policies]
\label{lem:slater}
Let $g$ satisfy the polynomial growth condition $g(\Delta)\le C(1+\Delta^{\ell})$.
\begin{enumerate}
\item[\textup{(a)}] Under GaW, for every $\bar K_{\max}>0$, some $\pi_s\in\Pi_G$ satisfies $\bar K_{\pi_s}<\bar K_{\max}$ and $\bar g_{\pi_s}<\infty$.
\item[\textup{(b)}] Under random arrivals, the same holds for every $\bar K_{\max}>\kappa_s\nu/(1+\nu\mathbb E[Y])$.
\end{enumerate}
\end{lemma}
\begin{proof}
For part (a), fix $d$ with $p_d\triangleq\mathbb P(Y\le d)>0$ and $w>0$. After each delivery, $\pi_{w,d}$ idles for $w$, samples, and resamples whenever the service age reaches $d$. The number $A$ of attempts per cycle is geometric with mean $1/p_d$, the cycle length satisfies $w\le L\le w+dA$, and the renewal-reward theorem gives $\bar K_{\pi_{w,d}}\le(\kappa_s+\kappa_p)/(p_dw)<\bar K_{\max}$ for large $w$. For part (b), $\pi_d$ preempts a busy arrival exactly when the service age exceeds $d$. With $\bar p_d\triangleq\mathbb P(Y>d)$, a cycle has one service start from idle and at most $\bar p_d/(1-\bar p_d)$ preemptions on average, and lasts at least $1/\nu+\mathbb E[\min\{Y,d\}]$ on average, so $\bar K_{\pi_d}\le(\kappa_s+\kappa_p\bar p_d/(1-\bar p_d))/(1/\nu+\mathbb E[\min\{Y,d\}])$, which tends to $\kappa_s\nu/(1+\nu\mathbb E[Y])$ as $d\to\infty$. Attaching to each attempt an independent $\mathrm{Exp}(\nu)$ clock that runs from age $d$ bounds the cycle length by an exponential idle period plus a geometric sum of light-tailed variables. In both parts, the cycle length $L$ thus has all moments, and so does the delivered service time $W$, which is at most $d$ under GaW and at most $d$ plus an $\mathrm{Exp}(\nu)$ clock under random arrivals. The running cost of a cycle is at most $L\,C(1+(W+L)^{\ell})$, with $W$ the previous delivered service time, so the Cauchy--Schwarz inequality bounds its expectation. The cycles form a one-dependent regenerative structure~\cite{sigman1990one}, so the limits defining $\bar g_{\pi_s}$ and $\bar K_{\pi_s}$ exist and are finite.
\end{proof}

\begin{proof}[Proof of Lemma~\ref{prop:duality}]
Write $V(b)\triangleq J_i(g;b)$ for $b\ge0$, with $V(b)=+\infty$ if no policy meets the budget $b$; since $g\ge0$, $V\ge0$. $V$ is nonincreasing in $b$.

For convexity, let $V(b_1),V(b_2)<\infty$, $q\in[0,1]$, and $\varepsilon>0$. Choose $\pi_j\in\Pi_i$ with $\bar K_{\pi_j}\le b_j$ and $\bar g_{\pi_j}\le V(b_j)+\varepsilon$ for $j=1,2$. Let $\pi_q$ use the independent randomization variable of Section~\ref{sec:model} at time zero to follow $\pi_1$ with probability $q$ and $\pi_2$ otherwise. Its expected cumulative costs are the $q$-mixtures of those of $\pi_1$ and $\pi_2$, so the limits defining $\bar g_{\pi_q}$ and $\bar K_{\pi_q}$ exist, with $\bar g_{\pi_q}=q\bar g_{\pi_1}+(1-q)\bar g_{\pi_2}$ and $\bar K_{\pi_q}=q\bar K_{\pi_1}+(1-q)\bar K_{\pi_2}\le qb_1+(1-q)b_2$. Hence $V(qb_1+(1-q)b_2)\le qV(b_1)+(1-q)V(b_2)+\varepsilon$, and letting $\varepsilon\downarrow0$ proves convexity.

The policy $\pi_s$ meets every budget $b>\bar K_{\pi_s}$, so $V\le\bar g_{\pi_s}<\infty$ on the open interval $(\bar K_{\pi_s},\infty)$, which contains $\bar K_{\max}$. A finite convex function on an open interval has finite one-sided derivatives $V'_-(\bar K_{\max})\le V'_+(\bar K_{\max})$, and every $s$ between them satisfies $V(b)\ge V(\bar K_{\max})+s(b-\bar K_{\max})$ for all $b\ge0$; monotonicity gives $s\le0$. Put $\lambda^\star\triangleq-s$. Every $\pi\in\Pi_i$ meets the budget $\bar K_\pi$, so
\begin{equation}
\bar g_\pi\ge V(\bar K_\pi)\ge J_i(g;\bar K_{\max})-\lambda^\star(\bar K_\pi-\bar K_{\max}).
\end{equation}
Taking the infimum of $\bar g_\pi+\lambda^\star\bar K_\pi$ over $\pi\in\Pi_i$ gives $\rho_i(g,\lambda^\star)-\lambda^\star\bar K_{\max}\ge J_i(g;\bar K_{\max})$, and weak duality gives the reverse inequality for every $\lambda\ge0$. This proves \eqref{eq:strong_duality}, with the maximum attained at $\lambda^\star$.
\end{proof}

\subsection{Lagrangian Envelope and Random Mixture}
\begin{lemma}[Lagrangian Envelope and Random Mixture]
\label{prop:kernel}
Suppose that $\Pr\{Y\le dt\}>0$, that $\Delta_{\mathrm{cut}}\ge2\,dt$ under random arrivals, and that some policy meets the budget. Let $\Pi_D$ be the finite set of deterministic stationary policies, and fix a running cost $g$. Then the following statements hold on the finite model.
\begin{enumerate}
\item[\textup{(i)}] $\rho_i(g,\lambda)=\min_{\pi\in\Pi_D}\{\bar g_\pi+\lambda\bar K_\pi\}$, so $\rho_i(g,\lambda)$ is concave, nondecreasing, and piecewise linear in $\lambda\ge0$.
\item[\textup{(ii)}] Suppose that $\pi_1,\pi_2\in\Pi_D$ both attain $\rho_i(g,\lambda)$ at some $\lambda\ge0$, with $\bar K_{\pi_1}>\bar K_{\max}\ge\bar K_{\pi_2}$, and let $q=(\bar K_{\max}-\bar K_{\pi_2})/(\bar K_{\pi_1}-\bar K_{\pi_2})$. Then the random mixture $\pi^\star$ that selects $\pi_1$ with probability $q$ at time zero and $\pi_2$ otherwise, and follows the selected policy thereafter, attains $J_i(g;\bar K_{\max})$, and its occupation measure satisfies \eqref{eq:budget_mixture}.
\end{enumerate}
\end{lemma}
\begin{proof}
We use the finite model of Section~\ref{subsec:tailor}. Under GaW, the finite model has busy-start states $y_i=i\,dt$, whose actions are preemption ages in $\Theta_h$, and idle states $t_j=j\,dt$, whose actions are grid waiting targets $z_j\ge t_j$; a stage runs from one decision to the next. Under random arrivals, a stage runs between consecutive arrivals, and its action is the binary preemption decision. Let $\mathcal S$ be the finite set of grid states and $\mathcal A(s)$ the finite action set at $s\in\mathcal S$. A stage that starts in state $s$ with action $a\in\mathcal A(s)$ lasts until the next decision, incurs expected running cost $c_g(s,a)$ and expected action cost $k(s,a)$, and has expected duration $\tau(s,a)$ and next-state distribution $P(\cdot\,|\,s,a)$. We write $c_\lambda(s,a)\triangleq c_g(s,a)+\lambda k(s,a)$ and $(Ph)(s,a)\triangleq\sum_{s'}P(s'\,|\,s,a)h(s')$ for functions $h$ on $\mathcal S$. Given $(s,a)$, the transition, duration, running-cost, and action-cost kernels of a stage do not depend on the actions taken in other states. An idle action may take zero time, but it enters a busy stage lasting at least $dt$, since preemption ages lie in $\Theta_h\subset(0,y_{\mathrm{cut}}]$ and service times are rounded up to the grid; eliminating these instantaneous transitions gives an equivalent SMDP with positive stage durations. Under any stationary policy, busy starts recur, and at each one service completes within the first step with probability $\Pr\{Y\le dt\}>0$, since a completion precedes a simultaneous arrival or preemption. This leads to the idle state with AoI $dt$ under GaW, and, followed by one arrival-free step and an arrival, to the busy-start state with AoI $2\,dt$ under random arrivals. Every state thus reaches a common state, so each stationary policy is unichain.

For part (i), the finite model has finite stage costs, positive finite mean stage durations, and a single recurrent class under every stationary policy, so it admits, for each $\lambda\ge0$, a deterministic stationary policy that minimizes $\bar g_\pi+\lambda\bar K_\pi$ over all policies~\cite[Ch.~11]{puterman1994markov}. Hence $\rho_i(g,\lambda)$ is the minimum of the finitely many affine functions $\lambda\mapsto\bar g_\pi+\lambda\bar K_\pi$, $\pi\in\Pi_D$, with nonnegative slopes $\bar K_\pi$, which makes it concave, nondecreasing, and piecewise linear in $\lambda$.

For part (ii), the selection at time zero is independent of the dynamics, so every expected finite-horizon cost of $\pi^\star$, and every expected occupation time, is the $q$-mixture of those of $\pi_1$ and $\pi_2$. Both stationary policies have convergent long-run averages, so $\bar g_{\pi^\star}=q\bar g_{\pi_1}+(1-q)\bar g_{\pi_2}$, $\bar K_{\pi^\star}=q\bar K_{\pi_1}+(1-q)\bar K_{\pi_2}=\bar K_{\max}$ by the choice of $q$, and $\mu_{\pi^\star}$ satisfies \eqref{eq:budget_mixture}. Because $\pi_1$ and $\pi_2$ both attain $\rho_i(g,\lambda)$, so does $\pi^\star$, and therefore $\bar g_{\pi^\star}=\rho_i(g,\lambda)-\lambda\bar K_{\max}$. Every policy $\pi$ that meets the budget satisfies $\bar g_\pi\ge\bar g_\pi+\lambda(\bar K_\pi-\bar K_{\max})\ge\rho_i(g,\lambda)-\lambda\bar K_{\max}=\bar g_{\pi^\star}$, so $\pi^\star$ attains $J_i(g;\bar K_{\max})$.
\end{proof}

The mixture $\pi^\star$ meets the budget in expectation, while its realized cost rate depends on the initial selection.

\textbf{Strong duality on the finite model.} On the finite model, let $d_i(g,\lambda;\bar K_{\max})\triangleq\rho_i(g,\lambda)-\lambda\bar K_{\max}=\min_{\pi\in\Pi_D}\{\bar g_\pi+\lambda(\bar K_\pi-\bar K_{\max})\}$, which is concave and piecewise linear by part (i). Since some policy meets the budget, part (i) applied to the action cost alone gives $\min_{\pi\in\Pi_D}\bar K_\pi\le\bar K_{\max}$, so the last affine piece of $d_i(g,\cdot\,;\bar K_{\max})$ has a nonpositive slope, so this function attains its maximum at some $\lambda^\star\ge0$. For a minimum of finitely many affine functions, the right derivative at a point equals the smallest slope among the pieces active there, and the left derivative equals the largest. If $\lambda^\star=0$, the nonpositive right derivative of $d_i(g,\cdot\,;\bar K_{\max})$ at zero yields a deterministic policy that attains $\rho_i(g,0)$ and meets the budget, and weak duality makes it optimal. If $\lambda^\star>0$, let $\Pi^\star\subseteq\Pi_D$ contain the policies that attain $\rho_i(g,\lambda^\star)$. The right and left derivatives of $d_i(g,\cdot\,;\bar K_{\max})$ at $\lambda^\star$ equal $\min_{\pi\in\Pi^\star}\bar K_\pi-\bar K_{\max}\le0$ and $\max_{\pi\in\Pi^\star}\bar K_\pi-\bar K_{\max}\ge0$. Hence either a policy in $\Pi^\star$ meets the budget exactly and is optimal by weak duality, or two policies in $\Pi^\star$ have action-cost rates on opposite sides of $\bar K_{\max}$ and part (ii) yields their optimal random mixture. In both cases $J_i(g;\bar K_{\max})=\max_{\lambda\ge0}d_i(g,\lambda;\bar K_{\max})$, so strong duality also holds on the finite model without linear programming.

\textbf{Dual lower bound.} For every function $h$ on $\mathcal S$, the semi-Markov extension of the \emph{Odoni bound}~\cite{odoni1969finding} reads
\begin{equation}\label{eq:bellman_lb}
\underline\rho(h)\triangleq\min_{s,a}\frac{c_\lambda(s,a)+(Ph)(s,a)-h(s)}{\tau(s,a)}\le\rho_i(g,\lambda).
\end{equation}
Under GaW, we apply it after eliminating instantaneous idle transitions, so the minimum runs over busy-start states. Indeed, the definition of $\underline\rho(h)$ gives $c_\lambda(s,a)+(Ph)(s,a)-h(s)\ge\underline\rho(h)\tau(s,a)$ for every state-action pair. Fix $\pi\in\Pi_D$ and let $\zeta_\pi$ be the invariant distribution of the embedded Markov chain of decision states under $\pi$, which is unique under the single-recurrent-class assumption. Averaging the inequality at $a=\pi(s)$ over $\zeta_\pi$, the terms in $h$ cancel by stationarity, so $\sum_s\zeta_\pi(s)c_\lambda(s,\pi(s))\ge\underline\rho(h)\sum_s\zeta_\pi(s)\tau(s,\pi(s))$. By the renewal-reward theorem, the ratio of these two sums equals $\bar g_\pi+\lambda\bar K_\pi$, so $\underline\rho(h)\le\bar g_\pi+\lambda\bar K_\pi$ for every $\pi\in\Pi_D$, and part (i) of Lemma~\ref{prop:kernel} yields \eqref{eq:bellman_lb}. If $h$ solves the optimality equation $h(s)=\min_a\{c_\lambda(s,a)-\rho_i(g,\lambda)\tau(s,a)+(Ph)(s,a)\}$, every ratio in \eqref{eq:bellman_lb} is at least $\rho_i(g,\lambda)$, with equality at a minimizing action, so the bound is tight. Evaluating \eqref{eq:bellman_lb} at the relative value function $h_\lambda$ that \textsc{Tailor} returns at each visited multiplier $\lambda\in\Lambda_{\mathrm{eval}}$ and applying weak duality give the lower bound that our implementation of Algorithm~\ref{alg:kernel} uses,
\begin{equation}\label{eq:kernel_lb}
\underline J(g)\triangleq\max_{\lambda\in\Lambda_{\mathrm{eval}}}\big\{\underline\rho(h_\lambda)-\lambda\bar K_{\max}\big\}\le J_i(g;\bar K_{\max}).
\end{equation}
The bracket $\underline J(g)\le J_i(g;\bar K_{\max})\le\overline J(g)$ therefore requires neither an exact multiplier search nor exact policy iteration, provided the reported bounds account for numerical errors.

\subsection{Proof of Lemma~\ref{lem:spending}}
\label{app:spending_proof}
\begin{proof}
(i) For $\lambda<\lambda'$, the optimality of $\pi_\lambda$ at $\lambda$ gives $\bar g_{\pi_\lambda}+\lambda\bar K_i(g,\lambda)\le\bar g_{\pi_{\lambda'}}+\lambda\bar K_i(g,\lambda')$, and the optimality of $\pi_{\lambda'}$ at $\lambda'$ gives the same inequality with the roles of $\lambda$ and $\lambda'$ swapped. Adding the two yields
\begin{equation}\label{eq:spend_monotone}
(\lambda'-\lambda)\big(\bar K_i(g,\lambda)-\bar K_i(g,\lambda')\big)\ge0,
\end{equation}
so $\bar K_i(g,\cdot)$ is nonincreasing. The finite model has finitely many deterministic stationary policies, so $\bar K_i(g,\cdot)$ takes finitely many values and, being monotone, is piecewise constant.

(ii) Since $\pi_\lambda$ attains $\rho_i(g,\lambda)$ and $\rho_i(g,\lambda')\le\bar g_{\pi_\lambda}+\lambda'\bar K_i(g,\lambda)$ for every $\lambda'\ge0$, \eqref{eq:dual_function} gives
\begin{equation}\label{eq:supergradient}
d_i(g,\lambda';\bar K_{\max})\le d_i(g,\lambda;\bar K_{\max})+(\lambda'-\lambda)\big(\bar K_i(g,\lambda)-\bar K_{\max}\big),
\end{equation}
so the line through $d_i(g,\lambda;\bar K_{\max})$ with slope $\bar K_i(g,\lambda)-\bar K_{\max}$ bounds $d_i(g,\cdot\,;\bar K_{\max})$ from above. By Lemma~\ref{prop:kernel}, $\rho_i(g,\cdot)$ is the minimum of finitely many affine functions on the finite model, so $d_i(g,\cdot\,;\bar K_{\max})$ is concave and piecewise linear.

(iii) If $\lambda<\lambda^\star$, then $\bar K_i(g,\lambda)>\bar K_{\max}$, and \eqref{eq:supergradient} gives $d_i(g,\lambda';\bar K_{\max})<d_i(g,\lambda;\bar K_{\max})$ for every $\lambda'<\lambda$, so $d_i(g,\cdot\,;\bar K_{\max})$ is nondecreasing on $[0,\lambda^\star)$. If $\lambda>\lambda^\star$, then $\bar K_i(g,\lambda)\le\bar K_{\max}$ by part~(i), and \eqref{eq:supergradient} gives $d_i(g,\lambda';\bar K_{\max})\le d_i(g,\lambda;\bar K_{\max})$ for every $\lambda'>\lambda$, so $d_i(g,\cdot\,;\bar K_{\max})$ is nonincreasing on $(\lambda^\star,\infty)$. Since $d_i(g,\cdot\,;\bar K_{\max})$ is continuous by part~(ii), $\lambda^\star$ maximizes it. The argument does not require $\bar K_i(g,\lambda^\star)\le\bar K_{\max}$, because the policy selected at a breakpoint may overspend. It also covers $\lambda^\star=0$, where only the nonincreasing part is needed.

For part~(iv), for $\lambda>0$, the optimality of $\pi_\lambda$ gives $\bar g_{\pi_\lambda}+\lambda\bar K_i(g,\lambda)\le\bar g_{\pi_s}+\lambda\bar K_{\pi_s}$, and $\bar g_{\pi_\lambda}\ge0$ because $g\ge0$, so $\bar K_i(g,\lambda)\le\bar K_{\pi_s}+\bar g_{\pi_s}/\lambda$. Since $\bar K_{\pi_s}<\bar K_{\max}$, this bound is at most $\bar K_{\max}$ for every $\lambda>0$ with $\lambda\ge\bar g_{\pi_s}/(\bar K_{\max}-\bar K_{\pi_s})$, so $\lambda^\star<\infty$.
\end{proof}

\subsection{Convergence of the Multiplier Bisection}
\label{app:bisection_proof}
\begin{lemma}[Convergence of the Multiplier Bisection]
\label{lem:bisection_gap}
Let $g\ge0$ with $\bar K_i(g,0)>\bar K_{\max}$, since otherwise Algorithm~\ref{alg:kernel} returns $\pi_0$ at $\lambda=0$, let $\pi_s\in\Pi_i$ be a policy with $\bar K_{\pi_s}<\bar K_{\max}$ and $\bar g_{\pi_s}<\infty$, and let $\bar\lambda\triangleq\max\{1,\bar g_{\pi_s}/(\bar K_{\max}-\bar K_{\pi_s})\}$. After the doubling stage and $n$ subsequent bisection steps, each of which solves \textsc{Tailor} once at the midpoint of $[\lambda_{\mathrm{lo}},\lambda_{\mathrm{hi}}]$, the mixture $\hat\pi$ in \eqref{eq:budget_mixture} meets the budget and satisfies
\begin{align}
0\le\bar g_{\hat\pi}-J_i(g;\bar K_{\max})&\le\bar g_{\hat\pi}-d_i(g,\lambda_{\mathrm{hi}};\bar K_{\max})\nonumber\\
&\le(\lambda_{\mathrm{hi}}-\lambda_{\mathrm{lo}})\big(\bar K_{\pi_{\mathrm{lo}}}-\bar K_{\pi_{\mathrm{hi}}}\big)\nonumber\\
&\le2^{-n}\,\bar\lambda\,\bar K_i(g,0).\label{eq:bisection_gap}
\end{align}
Hence, for any tolerance $\delta>0$, Algorithm~\ref{alg:kernel} reduces the gap between $\bar g_{\hat\pi}$ and the largest visited dual value to at most $\delta$ within $O(\log(1/\delta))$ solves of \textsc{Tailor} as $\delta\to0$.
\end{lemma}
\begin{proof}
As in the proof of part~(ii) of Lemma~\ref{prop:kernel}, the selection at time zero mixes all long-run averages linearly, so $\bar K_{\hat\pi}=q\bar K_{\pi_{\mathrm{lo}}}+(1-q)\bar K_{\pi_{\mathrm{hi}}}=\bar K_{\max}$ by the choice of $q\in[0,1)$, and $\bar g_{\hat\pi}=q\bar g_{\pi_{\mathrm{lo}}}+(1-q)\bar g_{\pi_{\mathrm{hi}}}$. Since $\pi_{\mathrm{hi}}$ attains $\rho_i(g,\lambda_{\mathrm{hi}})$ and $\bar K_{\pi_{\mathrm{hi}}}-\bar K_{\max}=-q(\bar K_{\pi_{\mathrm{lo}}}-\bar K_{\pi_{\mathrm{hi}}})$, we have $d_i(g,\lambda_{\mathrm{hi}};\bar K_{\max})=\bar g_{\pi_{\mathrm{hi}}}-q\lambda_{\mathrm{hi}}(\bar K_{\pi_{\mathrm{lo}}}-\bar K_{\pi_{\mathrm{hi}}})$, and therefore
\begin{equation}\label{eq:mixture_gap}
\bar g_{\hat\pi}-d_i(g,\lambda_{\mathrm{hi}};\bar K_{\max})=q\big[(\bar g_{\pi_{\mathrm{lo}}}+\lambda_{\mathrm{hi}}\bar K_{\pi_{\mathrm{lo}}})-(\bar g_{\pi_{\mathrm{hi}}}+\lambda_{\mathrm{hi}}\bar K_{\pi_{\mathrm{hi}}})\big].
\end{equation}
Since $\pi_{\mathrm{lo}}$ attains $\rho_i(g,\lambda_{\mathrm{lo}})$, we have $\bar g_{\pi_{\mathrm{lo}}}+\lambda_{\mathrm{lo}}\bar K_{\pi_{\mathrm{lo}}}\le\bar g_{\pi_{\mathrm{hi}}}+\lambda_{\mathrm{lo}}\bar K_{\pi_{\mathrm{hi}}}$. Substituting this inequality gives $\bar g_{\hat\pi}-d_i(g,\lambda_{\mathrm{hi}};\bar K_{\max})\le q(\lambda_{\mathrm{hi}}-\lambda_{\mathrm{lo}})(\bar K_{\pi_{\mathrm{lo}}}-\bar K_{\pi_{\mathrm{hi}}})$, which gives the third inequality in \eqref{eq:bisection_gap} because $q<1$ and $\bar K_{\pi_{\mathrm{lo}}}>\bar K_{\pi_{\mathrm{hi}}}$. The first inequality holds because $\hat\pi$ meets the budget, and the second is weak duality.

For the last inequality, the doubling starts from $[\lambda_{\mathrm{lo}},\lambda_{\mathrm{hi}}]=[0,1]$ and stops at the first $\lambda_{\mathrm{hi}}=2^k$ that meets the budget. The proof of Lemma~\ref{lem:spending}(iv) shows that every $\lambda\ge\bar\lambda$ meets the budget, so $k\le\lceil\log_2\bar\lambda\rceil$. If $k=0$, the initial bracket has width $1\le\bar\lambda$. If $k\ge1$, its width equals $\lambda_{\mathrm{lo}}=2^{k-1}$, and $\pi_{\lambda_{\mathrm{lo}}}$ overspends, so $\lambda_{\mathrm{lo}}<\bar\lambda$. After $n$ bisection steps, the width is therefore at most $2^{-n}\bar\lambda$. Moreover, $\bar K_{\pi_{\mathrm{lo}}}\le\bar K_i(g,0)$ by Lemma~\ref{lem:spending}(i), and $\bar K_{\pi_{\mathrm{hi}}}\ge0$ because action costs are nonnegative, which proves the last inequality in \eqref{eq:bisection_gap}.

For the stopping rule, \textsc{Tailor} returns exact relative value functions on the finite model, so the Odoni bound at each visited multiplier equals the dual value, and $\underline J\ge d_i(g,\lambda_{\mathrm{hi}};\bar K_{\max})$. The argument above gives $\overline J-d_i(g,\lambda_{\mathrm{hi}};\bar K_{\max})\le2^{-n}\bar\lambda\bar K_i(g,0)$, so the loop of Algorithm~\ref{alg:kernel} ends after at most $\lceil\log_2(\bar\lambda\bar K_i(g,0)/\delta)\rceil^+$ bisection steps. Adding the solves at $\lambda=0$ and $\lambda=1$ and at most $\lceil\log_2\bar\lambda\rceil$ doubling solves gives at most $2+\lceil\log_2\bar\lambda\rceil+\lceil\log_2(\bar\lambda\bar K_i(g,0)/\delta)\rceil^+$ solves of \textsc{Tailor}, where $\lceil x\rceil^+\triangleq\max\{0,\lceil x\rceil\}$, which is $O(\log(1/\delta))$.

\end{proof}

\subsection{Proof of Lemma~\ref{prop:density}}
\label{app:density_proof}
Let $\bar r_R=\nu$ and $\bar r_G=\bar K_{\max}/\kappa_s$, where GaW requires $\kappa_s>0$ and $\bar K_{\max}>0$. The lemma also holds on the finite model of Section~\ref{subsec:tailor}, whose AoI drops only at deliveries, otherwise advances by the elapsed time, and has delivery rate at most $\bar r_i$. Here the occupation measure follows the unit-rate AoI trajectory between grid events, not the quantized decision states. Under GaW, this holds by construction: a stage advances the AoI by its own duration, a delivery resets it to the quantized service time, and the far-field closure continues the same unit-rate evolution. Under random arrivals, arrival epochs fall off the grid, so the numerical model must round them upward; downward rounding at a non-delivery event would add artificial downward jumps that inflate the time spent in an age window and void the bound.
\begin{proof}
Let $N_c(T)$ count deliveries in $[0,T]$. Between consecutive deliveries, the AoI grows at unit rate and spends at most $b-a$ time units in $(a,b]$, so $\int_0^T\mathbf 1_{\{a<\Delta(t)\le b\}}\,dt\le(b-a)[N_c(T)+1]$. Under random arrivals, every delivery consumes one arrival, so $N_c(T)$ is at most the number of arrivals in $[0,T]$ plus one for an update already in service at time zero, and $\limsup_T\mathbb E[N_c(T)]/T\le\nu$. Under GaW, every completed busy period is preceded by an idle sample of cost $\kappa_s$, apart from a possible initial busy period, so $\kappa_s[N_c(T)-1]$ never exceeds the action cost incurred in $[0,T]$ and budget feasibility gives $\limsup_T\mathbb E[N_c(T)]/T\le\bar K_{\max}/\kappa_s$. Taking expectations in the occupation inequality, dividing by $T$, and passing to the limit proves \eqref{eq:search_density_bound}; a monotone-class argument extends it to $\mu_\pi(B)\le\bar r_i\,\mathrm{Leb}(B)$ for every Borel set $B\subseteq(0,\infty)$, and together with $\mu_\pi(\{0\})=0$, the Radon--Nikodym theorem gives a Lebesgue density bounded by $\bar r_i$. Hence $\mathbb P(\tilde\Delta_\pi\le x)\le\bar r_ix$ for $x\ge0$, and $\operatorname{VaR}_\alpha(\tilde\Delta_\pi)\ge\alpha/\bar r_i$.
\end{proof}

\subsection{Proof of Theorem~\ref{prop:cvar_guarantee}}
\label{app:cvar_proof}
\begin{proof}
\textbf{Curvature and search range.} We use the primal form $H_i(\eta)=\inf_\pi\{\eta+\mathbb E[(\tilde\Delta_\pi-\eta)^+]/(1-\alpha)\}$, the infimum being over budget-feasible policies; by cost consistency, it agrees with \eqref{eq:cvar_outer}. With $F_\pi$ as in Section~\ref{subsec:cvar}, we have $F_\pi'(\eta)=1-\mathbb P(\tilde\Delta_\pi>\eta)/(1-\alpha)\in[-\frac{\alpha}{1-\alpha},1]$, and the density bound makes $F_\pi'$ absolutely continuous with $F_\pi''\le M$ almost everywhere. The two-sided increment bounds hold for every feasible $\pi$ and pass to $H_i=\inf_\pi F_\pi$, the upper one by applying it to an $\varepsilon$-minimizer at $\eta_1$ and letting $\varepsilon\downarrow0$:
\begin{equation}\label{eq:H_slope}
-\frac{\alpha}{1-\alpha}\le\frac{H_i(\eta_2)-H_i(\eta_1)}{\eta_2-\eta_1}\le1,\qquad 0\le\eta_1<\eta_2.
\end{equation}
Each $F_\pi(\eta)-M\eta^2/2$ is concave, and so is their pointwise infimum $H_i(\eta)-M\eta^2/2$. By Lemma~\ref{prop:density}, $\operatorname{VaR}_\alpha(\tilde\Delta_\pi)\ge\alpha/\bar r_i$. Since each $F_\pi$ is convex and attains its minimum at $\operatorname{VaR}_\alpha(\tilde\Delta_\pi)$, it is nonincreasing on $[0,\alpha/\bar r_i]$, and so is $H_i$; hence $H_i(\eta)\ge H_i(\alpha/\bar r_i)$ for $0\le\eta\le\alpha/\bar r_i$, where $\eta\ge0$ as in \eqref{eq:cvar_outer}, and $U\ge\operatorname{VaR}_\alpha(\tilde\Delta_{\pi_{\mathrm{ref}}})\ge\alpha/\bar r_i$. Also, $H_i(\eta)\ge\eta$, so values above $U$ need not be searched, and the continuity from \eqref{eq:H_slope} handles the endpoint. Hence, for every finite $U\ge C^\star_{\bar K_{\max},\alpha,i}$, the infimum in \eqref{eq:cvar_outer} can be restricted to $[\alpha/\bar r_i,U]$.

\textbf{Interval lower bounds.} For $a<b$, put $d=b-a$, $L_a=\underline H(a)$, $L_b=\underline H(b)$, $c=\alpha/(1-\alpha)$, and $Q_{a,b}(s)=L_a+(L_b-L_a)s/d-Ms(d-s)/2$ for $0\le s\le d$. The slope bounds \eqref{eq:H_slope} and the concavity of $H_i(\eta)-M\eta^2/2$ give the valid bound
\begin{equation}\label{eq:interval_lb_full}
\mathrm{LB}(a,b)=\max\Big\{a,\ \frac{L_a+cL_b-cd}{1+c},\ \min_{0\le s\le d}Q_{a,b}(s)\Big\}.
\end{equation}
The first term uses $H_i(\eta)\ge\eta$. For the second, \eqref{eq:H_slope} gives $H_i(a+s)\ge L_a-cs$ and $H_i(a+s)\ge L_b-(d-s)$, and the convex combination with weights $1/(1+c)$ and $c/(1+c)$ eliminates $s$; the term is valid even when the two lines cross outside the interval. The third follows from this concavity by replacing the endpoint values by their lower bounds, which preserves the bound because both endpoint weights are nonnegative. The minimizer of $Q_{a,b}$ is $s^\star=\operatorname{clip}\big(d/2-(L_b-L_a)/(Md),0,d\big)$, and $s(d-s)\le d^2/4$ gives
\begin{equation}\label{eq:interval_bound_summary}
\mathrm{LB}(a,b)\ge\min\{\underline H(a),\underline H(b)\}-Md^2/8.
\end{equation}

\textbf{Guarantee and complexity.} At every evaluated threshold $\eta$, Algorithm~\ref{alg:kernel} returns a budget-feasible policy $\pi_\eta$ and bounds with $\underline J\le J_i((\Delta-\eta)^+;\bar K_{\max})\le\bar g_{\pi_\eta}\le\overline J$, so $\operatorname{CVaR}_\alpha(\tilde\Delta_{\pi_\eta})\le\eta+\bar g_{\pi_\eta}/(1-\alpha)\le\overline H(\eta)$. The algorithm updates the incumbent with every $\overline H(\eta)$, so $C^\star_{\bar K_{\max},\alpha,i}\le C_{\mathrm{up}}\le\overline H(\eta)$ at every evaluated point. The active intervals cover the search range and each interval bound is valid, so $C_{\mathrm{lo}}=\min_{[a,b]}\mathrm{LB}(a,b)$ is a global lower bound. Since the saved policy has CVaR at most $C_{\mathrm{up}}$, its gap never exceeds $C_{\mathrm{up}}-C_{\mathrm{lo}}$. For the evaluation count, Algorithm~\ref{alg:cvar_search} calls Algorithm~\ref{alg:kernel} with tolerance $(1-\alpha)\gamma\epsilon$, so Lemma~\ref{lem:bisection_gap} gives $\overline H(\eta)-\underline H(\eta)\le\gamma\epsilon$ at every evaluated threshold. On an interval of length $d$, \eqref{eq:interval_bound_summary} gives $\mathrm{LB}(a,b)\ge C_{\mathrm{up}}-\gamma\epsilon-Md^2/8$, so an interval selected for subdivision while the stopping test fails must have $d>d_\epsilon\triangleq\sqrt{8(1-\gamma)\epsilon/M}$, and every child it creates is longer than $d_\epsilon/2$. Let $n_0$ be the number of initial thresholds. Each bisection adds one evaluation and one leaf interval, and at termination the leaf intervals created by bisections are pairwise disjoint and longer than $d_\epsilon/2$, so there are fewer than $2(U-\alpha/\bar r_i)/d_\epsilon$ bisections and at most $n_0+(U-\alpha/\bar r_i)\sqrt{M/(2(1-\gamma)\epsilon)}$ calls of Algorithm~\ref{alg:kernel}. This proves the stated bound. For the total number of \textsc{Tailor} solves, assume a policy $\pi_s$ with $\bar K_{\pi_s}<\bar K_{\max}$ and finite average AoI, independent of $\eta$. Since $g_\eta\le\Delta$, its running cost is at most its average AoI, so $\bar\lambda$ in Lemma~\ref{lem:bisection_gap} is uniform in $\eta$. Moreover, $\bar K_i(g_\eta,0)$ is at most the largest average action cost over the finitely many deterministic stationary policies of the finite model. Each call of Algorithm~\ref{alg:kernel} therefore uses $O(\log(1/\epsilon))$ solves of \textsc{Tailor}, and multiplying by the number of calls proves the total. Reference endpoints and reported risks are evaluated exactly on the finite model.
\end{proof}

\end{document}